\documentclass[10pt,a4paper]{article}
\usepackage{makeidx}
\usepackage{amssymb}
\usepackage{amsmath}
\usepackage{graphicx}
\usepackage{graphics}
\usepackage{times}
\usepackage[T1]{fontenc}
\usepackage[thinspace,thinqspace,squaren]{SIunits}
\usepackage{graphicx}
\usepackage{amsthm}
\usepackage{color}
\usepackage{hyperref}
\usepackage{multirow}

\usepackage{ctable}
\allowdisplaybreaks

\usepackage{algorithm}
\usepackage{algorithmic}

\onecolumn

\begin{document}
	\date{}

\title{Control of a Rigid Wing Pumping Airborne Wind Energy System in all Operational Phases}

\author{Davide~Todeschini  and 
	Lorenzo~Fagiano and Claudio~Micheli and Aldo~Cattano
	\thanks{D. Todeschini and L. Fagiano are with the Dipartimento di Elettronica, Informazione e Bioingegneria, Politecnico di Milano, Piazza Leonardo da Vinci
		32, 20133 Milano, Italy. Claudio Micheli is now with Auterion Inc., Giessh\"{u}belstrasse 40,
		8045 Z\"{u}rich,	Switzerland. Aldo Cattano is with Skypull SA, Via alla Stampa 49, 6967 Lugano, Switzerland. Corresponding author L. Fagiano \{lorenzo.fagiano \}@polimi.it. \newline This is a preprint of a  paper submitted to Control Engineering Practice.}}

\maketitle

\begin{abstract}
	The control design of an airborne wind energy system with rigid aircraft, vertical take-off and landing, and pumping operation is described. A hierarchical control structure is implemented, in order to address all operational phases: take-off, transition to power generation, pumping energy generation cycles, transition to hovering, and landing. Control design at all hierarchical levels is described. The design approach is conceived and developed with real-world applicability as main driver. Aircraft design considerations in light of system maneuverability are presented, too, as well as three possible alternative strategies for the retraction phase of the pumping cycle. The automatic control approach is assessed in simulation with a realistic model of the overall system, and the results yield a comparison among the three retraction strategies, clearly indicating the most efficient one.  The presented results allow one to simulate the dynamical behavior of an AWE system in all operational phases, enabling further studies on all-round system automation, towards fully autonomous and reliable operation.
\end{abstract}



\section{Introduction}\label{S:introduction}

An Airborne Wind Energy (AWE) system converts wind energy into electricity with an autonomous aircraft that carries out periodic trajectories in the wind flow \cite{Cherubini2015,Schmehl2018} and is tethered to a ground station.
Deemed a potential game-changing solution \cite{IRENA16}, AWE is attracting the attention of researchers, entrepreneurs, and policy makers alike \cite{Study_AWES,WATSON2019109270}, with the promise of producing large amounts of cost-competitive electricity and with wide applicability worldwide \cite{FaMP09,Archer2009,Archer2014,BECHTLE20191103}. Today, AWE is the umbrella name for a series of technologies under development, which can be classified according to different criteria, such as the  operating principle (drag power or pumping operation), aircraft type (rigid or flexible), or take-off and landing method (vertical or linear), see, e.g., \cite{Schmehl2018,Study_AWES} for an overview and \cite{VanderLind2013,BAUER2018290,Erhard2015,Zgraggen2016,LICITRA201815,LKBWRD19,Fechner2015,Heilmann2013,ch26-RuSo14,Stuyts2015,FNRSO18,VCSPW18,Twingtec2018,Todeschini2019,RSOH19} for contributions on specific aspects.\\
No fully autonomous AWE system has been commercialized so far: many groups in industry and academia are working to develop the technologies. Fully autonomous and reliable operation is currently one of the major research and development priorities \cite{Study_AWES}. An essential component to reach this goal is the automatic control system that shall operate the aircraft and the ground station in all the different system and environment conditions, and in all phases: take-off, power generation, and landing. Such a control system features a hierarchical topology, with control functions at different layers distributed across the various subsystems. In the scientific literature, most contributions focus on control design for the power generation phase, mainly with flexible wings (see, e.g., \cite{Erhard2015,Zgraggen2016,Fechner2015}) but also with rigid ones
\cite{HaOD18,LKBWRD19,RSOH19}. A few works deal with the control aspects of take-off and/or landing phases \cite{Bont10,Fagianoa,FNRSO18}. To the best of the authors' knowledge, no contribution so far has described the control of an AWES in \textit{all} operational phases, allowing one to simulate the dynamical system behavior from take-off, to power generation, to landing, including all transition phases, and eventually to implement the control approach on real-world systems.\\
This paper provides its main contribution in this direction, by describing a design approach for an automatic control logic able to drive an AWE system in all operational phases. The considered concept employs a rigid aircraft with vertical take-off and landing (VTOL) capability, single tether, and pumping operation, in particular the one with boxed-wing design developed by Skypull SA \cite{Skypull}. We adopt a realistic model of the whole system, and design the feedback controllers at all hierarchical levels, from low-level attitude control up to the supervisory state machine that defines the phase transitions, to realize fully autonomous operation in normal (i.e., non-faulty) conditions. In the same spirit as \cite{RSOH19}, the control system design is approached with simplicity and effectiveness in mind, making it highly suited for implementation on a real prototype. Moreover, this article delivers two additional contributions: 1) a study on the links between aircraft/control design and its maneuverability, exploring the minimum steering radius for a given design as a function of the wind speed, and 2) three possible strategies for the retraction (or reentry) phase of the pumping cycle: a ``free-flight'' one, where the aircraft glides upwind with very low force on the tether, and two alternatives with taut tether and different flight trajectories. The automatic control system is assessed via numerical simulations, showing the good performance of the approach and allowing us to compare the three reentry alternatives. The comparison clearly confirms that the free-flight strategy achieves the best conversion efficiency of the pumping cycle, with about 70\% ratio between cycle power and traction phase power. A very preliminary study connected to this paper is \cite{Todeschini2019}, where we considered the control in all phases of an untethered aircraft, hence with a simpler model, without the system and control design aspects for tethered flight and power generation, and without the study on different reentry strategies.\\
The paper is structured as follows. In Section \ref{S:sys_descr}, a concise description of the system under study and its operational phases is given. Section \ref{S:steering} presents the analysis of aircraft design vs. maneuverability. Section \ref{S:model} describes the dynamical system model, divided into its main subsystems: the drone, the tether, and the ground station. Section \ref{S:control} presents the automatic control strategy with the different reentry alternatives. Section \ref{S:results}  describes simulation results and the comparison among the reentry strategies. Finally, Section \ref{S:conclusions} contains concluding remarks and points out the next research and development steps.

\section{System description and operating principle}\label{S:sys_descr}
\subsection{System description}\label{SS:system_description}
\begin{figure}[hbt!]
	\centering
	\includegraphics[width=\columnwidth]{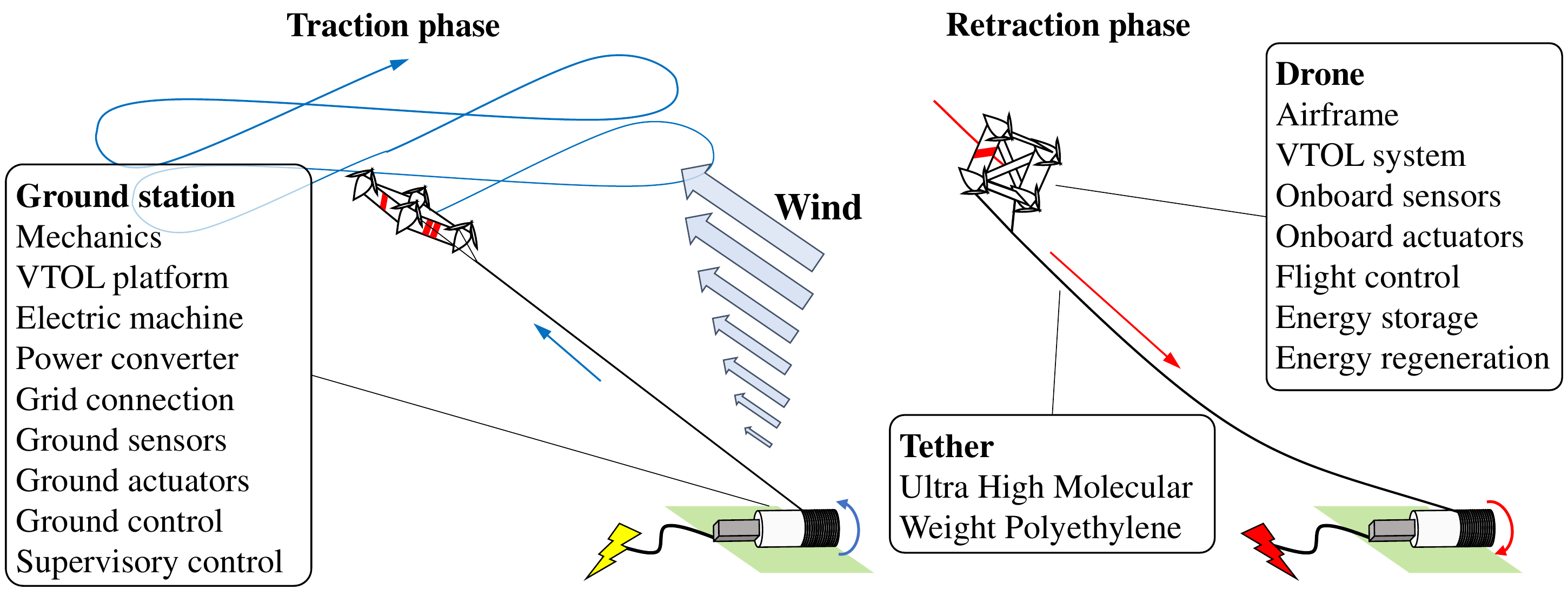}
	\caption{Conceptual sketch of the considered AWE system with its main components and of pumping operation.}\label{F:system_sketch}
\end{figure}

\begin{figure}
	\centering
	\centerline{\includegraphics[width=.6\columnwidth]{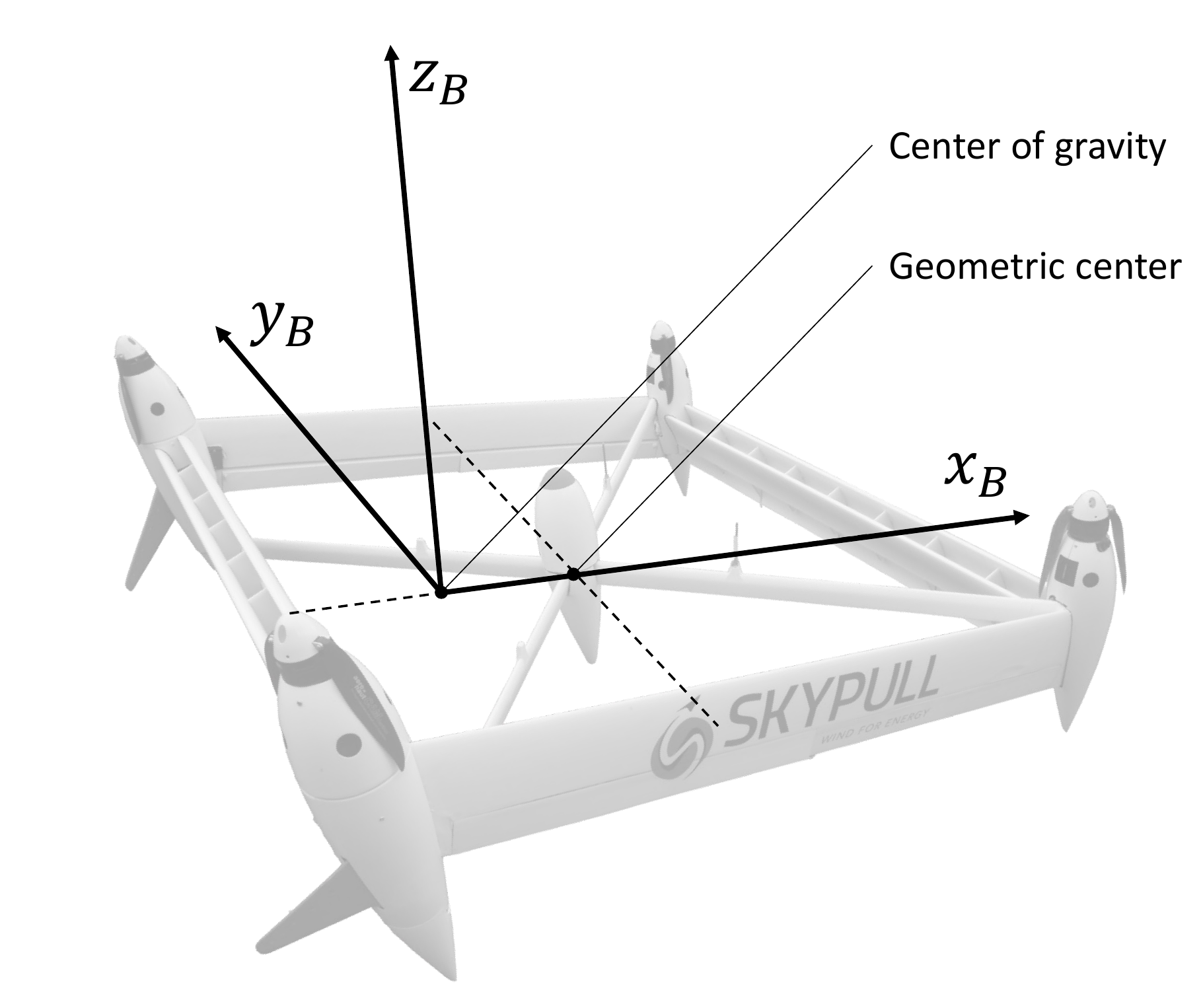}}
	\caption{Position of the center of gravity and geometric center of the drone, and adopted body reference frame $B$.}
	\label{F:reference}
\end{figure}

A sketch of the system considered in this paper is presented in Figure \ref{F:system_sketch}. It employs a tethered rigid wing aircraft (also referred to as ``drone'' in the remainder) with a single tether and ground-based power conversion via pumping cycle operation. The drone has VTOL capability thanks to onboard propellers.
One of the main peculiarities of this system with respect to the other concepts currently under development is the boxed-wing design of the drone with a propeller at each corner, see Figure \ref{F:reference}. Due to the distribution of masses and the positioning of batteries in the lower corners of the frame, the geometric center and the center of gravity are different (see Figure \ref{F:reference}).
This allows one to identify a lower wing, the one that is closer to the center of gravity, and an upper one (i.e. the opposite one), and consequently  a left and right wing by looking at the drone from behind.\\
\noindent On each lateral wing there are two discrete control surfaces, while the upper and lower wings feature three discrete control surfaces each. These surfaces can be actuated either symmetrically, on a single wing or on two opposite wings, leading to a symmetric change of aerodynamic coefficients (hence of lift and drag forces), or in opposite directions. In the second case, the effect is to generate  turning moments. All the possible combinations of movements of control surfaces lead to a rather high control authority when operating in dynamic flight mode. In addition, the four propellers act as in standard quad-copters in hovering mode, and they can further contribute to turning moments also in dynamic flight mode, besides providing the forward thrust to keep cruise velocity. In this report, we assume that the control system can inject turning moments around the axes of the body coordinate system $B=(x_B,y_B,z_B)$ depicted in Figure \ref{F:reference}, plus a thrust force in the $z_B$ direction. Therefore, we have four control inputs (three turning moments and the thrust force). The translation from turning moments to individual surface/propeller commands is managed by a low-level logic that aims to provide the wanted action with minimal energy consumption. When operating in tethered mode in presence of strong-enough wind, the propellers are not used to generate thrust force, rather the forward apparent speed is provided by the well-known phenomena of crosswind kite power \cite{Loyd1980},\cite{FaMi12}.\\
Regarding the tether, this is made of Ultra-High Molecular Weight Polyethylene (UHMWPE), as in most pumping AWE systems. Finally, the ground station features a winch connected to an electric machine that serves both as generator and as motor during pumping operation, together with all the required subsystems (power electronics, energy storage, mechanical frame, tether spooling system, ground control system, ground sensors, etc.).

\subsection{Operating principle}\label{SS:operating_principle}
In normal (i.e. non-faulty) conditions, the working principle of this class of system features the following phases:
\begin{itemize}
	\item {\bf Vertical take-off}. When the wind conditions at the target operating altitude 
	 are suitable to generate energy (cut-in wind speed), the system takes-off vertically using the on-board propellers, in hovering mode. This is the typical flight condition of quad-rotors, with the addition of the tether. The propellers sustain the weight of the vehicle and control the attitude.
		\item {\bf Transition from hovering to power generation}. In this phase, the drone must quickly speed up and rotate to achieve the attitude and reference speed that can sustain its weight during the power generation phase.
	\item {\bf Power generation}. The system transitions from multicopter mode to dynamic flight and enters into power generation mode. This exploits the so-called ``pumping operation'', composed of two phases (see Figure \ref{F:system_sketch}): in the traction phase the drone flies fast in crosswind patterns (i.e. roughly perpendicular to the wind flow), like a steerable kite, and the tether is reeled-out under high force, generating energy, while in the retraction phase the drone glides towards the ground station and the tether is quickly reeled-in under very low force, spending about 10\% of the energy previously generated. Two transition phases link the traction and retraction ones to achieve a repetitive power generation cycle. During power generation, the aircraft is kept airborne by large aerodynamic forces, the onboard propellers are therefore not employed, or they are used as generators to recharge the batteries and supply power to the onboard electronics and actuators. This is the typical flight condition of an airplane, with the  addition of the tether.
	\item {\bf Transition from flight to hovering}. In this phase, the drone must slow down and rotate with propellers pointing up, to achieve a stationary hovering condition.
	\item 	{\bf Vertical landing}. when the wind is too weak to generate energy, or too strong to operate the system safely, the drone carries out a landing manoeuvre composed of a first gliding part up to about 50 m of altitude, followed by a controlled vertical landing, after transitioning from dynamic flight back into multicopter mode.
\end{itemize}
\noindent The main overall control objective is to obtain a fully autonomous flight cycle from take-off to landing, going through all the phases described above.\\
The model and control algorithms presented in this paper allow one to realize such an all-round operation.

\subsection{Alternative reentry strategies}\label{SS:cycle}
In order to maximize the efficiency of the pumping cycle, it is important analyze in detail the reentry phase and select the best strategy among different possible ones. One of the contributions of this paper is indeed to compare three different alternatives and their advantages and disadvantages. It is appropriate to describe qualitatively these three alternative strategies at this point, while in the next chapters we will point out the corresponding specific control solutions. In the remainder, we adopt the term ``free-flight'' to indicate a flight mode with slack tether, as opposed to ``taut-tether'' flight. Moreover, it is assumed that the traction phase is carried out with figure-of-eight flight patterns in crosswind conditions.

\subsubsection{Free-flight reentry}
With this strategy, the reentry starts with the drone transitioning from taut-tether flight into free-flight. The ground station controller regulates the tether speed in order to limit the pulling force at very small values, thus limiting interference with the drone control system that aims to stabilize the drone in a free-flight condition. Then, the drone starts to glide upwind towards the ground station at controlled speed. To facilitate the restart of the next traction phase, in which the drone enters a figure of eight path, the gliding trajectory is carried out pointing to one side of the ground station with respect to the wind direction. When the drone is sufficiently close to the ground station, its control systems has to cooperate with the ground station's one in order to transition again from free-flight to tethered flight. This is the most critical part of the trajectory and requires a very accurate coordination of on-board and on-ground controllers. 


\subsubsection{Complete rotation around the ground station}
The second reentry strategy is a lateral reentry with a complete rotation around the ground station. In this case, the tether is maintained taut at all times, and the ground station controller has to reel-in at constant speed. Note that the drone makes a spiraling reentry, rotating around the ground station. Furthermore, to maximize the cycle efficiency, the reentry controller reference velocity has to be adequately calibrated, in order to have no high peaks of required power and at the same time guarantee a sufficiently fast tether reel in.

\subsubsection{Climb and descend reentry}
In the third reentry strategy, the drone continues to follow a figure of eight path, but at an higher altitude than the one used in the traction phase. Doing this maneuver, the elevation angle increases and the relative wind seen by the drone decreases. As a consequence, the force needed to reel in the tether can be reduced. Also in this case, the reference tether velocity has to be suitably calibrated to optimize the cycle performance.


\section{Steering authority analysis and drone design considerations}\label{S:steering}
Steering authority during tethered flight is a crucial aspect for an AWE system, because of the particular trajectories to be achieved. We term steering authority of the tethered system the minimum turning radius that it can achieve while flying with taut tether under high lift force and apparent speed. The smaller such a steering radius, the higher the steering authority. The latter is affected by different aspects, including the steering strategy which, for the considered box-wing design, can be either  ``roll-based'' or ``yaw-based''.\\
In our research, we went through an initial design phase in which we analyzed these two options, and more in general the link between system design and steering authority, until eventually opting for a yaw-based strategy. In this section, we present the key points of our analysis, since they motivate subsequent control design choices and they can be useful also for other AWE systems' developments. In particular, we analyze the steering authority in two different conditions: 1) trajectories parallel to ground, and 2) trajectories perpendicular to ground. These are indeed two extreme cases of all the situations that can occur in tethered crosswind flight. In the analysis, we assume for the sake of simplicity that the apparent speed magnitude coincides with the drone's velocity vector magnitude (i.e. no absolute wind speed is present). The results hold qualitatively also in presence of wind.
\subsection{Trajectories parallel to ground}
The centrifugal force during a turn at constant radius $r$ can be expressed as:
\begin{equation}\label{centrifugal_force}
F_{centrifugal}=m\frac{v^2}{r}
\end{equation}
where $m$ is the drone mass and $v$ is its speed. The centrifugal force has to be counterbalanced by a centripetal one in order to turn at constant radius $r$. The generation mechanism of the centripetal force depends on the considered steering strategy, as described below.

\subsubsection{Roll-based steering strategy} 

\begin{figure}[hbt!]
	\centering
	\includegraphics[width=1\columnwidth, keepaspectratio]{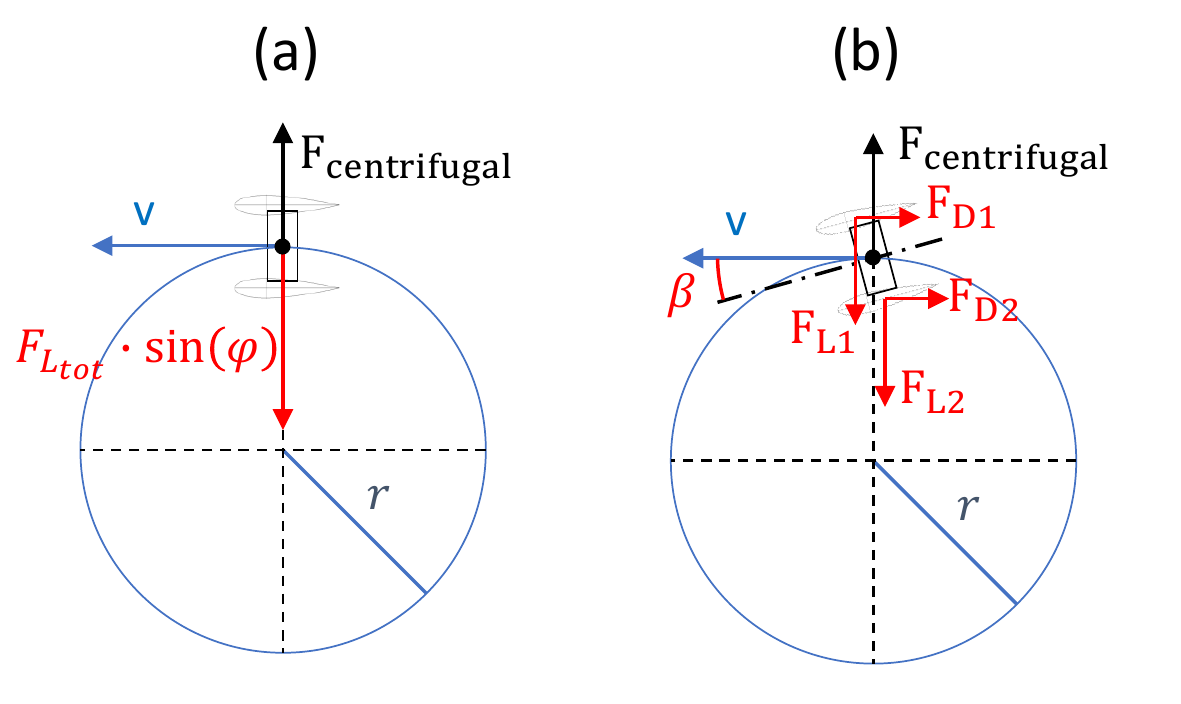}
	\caption{Sketch of steering using (a) roll-based or (b) yaw-based strategies.}
	\label{steering_roll}
\end{figure}

With roll-based steering, the centripetal force is the projection of the lift force on the plane where the drone's trajectory lies (see Figure \ref{steering_roll}-(a)):
\begin{equation}\label{F:centripetal_roll}
F_{centripetal}=F_{L}\sin{\varphi}
\end{equation}
where $F_{L}$ is the total lift force (generated by upper and lower aerodynamic surfaces) and $\varphi$ is the roll angle.
The lift force is computed as:
\begin{equation}\label{lift_force}
F_{L}=\frac{1}{2}\rho S c_{L}(\alpha,\beta)v^2
\end{equation}
where $\rho$ the air density, $S$ is the effective wing surface, $c_L$ the lift coefficient, and $\alpha,\beta$ the angle of attack and side-slip angle, respectively. The aerodynamic coefficients employed in our study are reported in Section \ref{S:results} (Figure \ref{F:aero_coeff}).\\
Equating \eqref{centrifugal_force} and \eqref{F:centripetal_roll} and inserting
 \eqref{lift_force} into \eqref{F:centripetal_roll} we have
\begin{equation}\label{roll_equation2}
 \frac{1}{2}\rho S c_{L}(\alpha,\beta)v^2\sin{\varphi}=m\frac{v^2}{r}
\end{equation}
solving for the turning radius, we finally obtain:
\begin{equation}\label{roll_radius}
r= \frac{2m}{\rho (S_{up}+S_{down}) c_{L}(\alpha)\sin{\varphi}}
\end{equation}
From this equation it can be noted that, to increase the steering authority using a roll-based strategy, one of the following measures can be adopted, compatibly with other design and control requirements:
\begin{itemize}
	\item Reduce the drone mass $m$;
	\item Increase the upper and lower wings surfaces $S_{up}, S_{down}$;
	\item Increase the lift coefficient $c_{L}$ (e.g. higher angle of attack $\alpha$);
	\item Increase the roll angle $\varphi$.
\end{itemize}

\subsubsection{Yaw-based steering strategy}

In yaw-based  steering (see Figure \ref{steering_roll}-(b)), the centripetal force is provided by the total lift force $F_{L_{lateral}}$ of the lateral wings. In this case, the force balance yields:
\begin{equation}\label{yaw_equation}
\begin{array}{ccc}
F_{centripetal}=F_{centrifugal} \\
\frac{1}{2}\rho (S_{right}+S_{left})c_{L_{lateral}}(\beta)v^2 =m\frac{v^2}{r}
\end{array} 
\end{equation}
where $S_{right}, S_{left}$ are, respectively, the surface of the left and right wings, and $c_{L_{lateral}}(\beta)$ their lift coefficient. The latter is a function of the side-slip angle $\beta$, which for the lateral wings corresponds to the angle of attack. The turning radius is thus given by:
solving for turning radius
\begin{equation}\label{yaw_radius}
r= \frac{2m}{\rho (S_{right}+S_{left}) c_{L_{lateral}}(\beta)}
\end{equation}
Thus, to increase the steering authority with a yaw-based strategy, one of the following measures can be adopted:
\begin{itemize}
	\item Reduce the drone mass $m$;
	\item Increase the lateral wings' surfaces $S_{right}, S_{left}$;
	\item Increase the lift coefficient of the lateral wings, $c_{L_{lateral}}$;
	\item Increase the side slip angle $\beta$ (without exceeding its limits for a stable flight).
\end{itemize}
\subsection{Trajectories perpendicular to ground}
In the previous section, results for trajectories parallel to ground have been presented. However, in case of up-loops or down-loops, the steering ability is affected also by the gravity force $F_g$ and its orientation during the path. In a part of the path, $F_g$ has a centripetal component, which helps during the turns. In the opposite part of the path, $F_g$ has a centrifugal component, which makes the turns more difficult. Thus, during trajectories not parallel to ground, steering ability (and minimum turning radius) is bounded between 2 extreme values, as described below.

\subsubsection{Top of trajectory}

\begin{figure}
\centering
\includegraphics[width=.8\columnwidth, keepaspectratio]{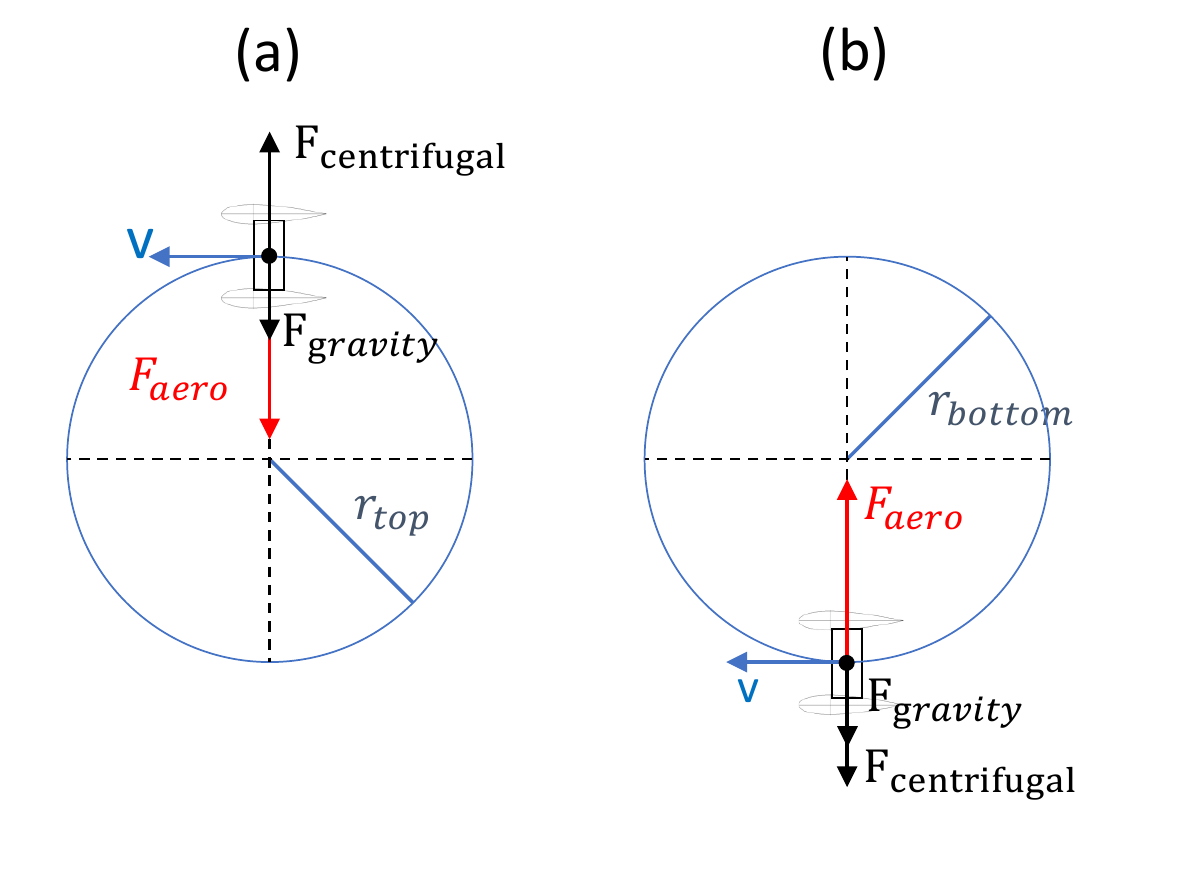}
\caption{Sketch of turning maneuver with path perpendicular to ground: top of trajectory (a) and bottom of trajectory (b).}
\label{steering_top}
\end{figure}

Referring to Figure \ref{steering_top}-(a), we have:
\begin{equation}
F_{centrifugal}=F_{centripetal}+mg
\end{equation}
solving for the turning radius
\begin{equation}\label{top_radius}
r_{top}=\frac{mv^2}{F_{centripetal}+mg}
\end{equation}
where $F_{centripetal}$ is provided by either one of the two strategies presented in the previous section.
From eq. \eqref{top_radius} it can be noted that, in the upper part of the trajectory, the gravity force is a centripetal one, helping the drone to turn. Moreover, the turning radius increases with larger speed values.
\subsubsection{Bottom of trajectory}


Referring now to Figure \ref{steering_top}-(b), we have:
\begin{equation}
F_{centrifugal}=F_{centripetal}-mg
\end{equation}
and solving for the turning radius we obtain
\begin{equation}\label{bottom_radius}
r_{bottom}=\frac{mv^2}{F_{centripetal}-mg}
\end{equation}
from which it can be noted that, in the bottom part of the trajectory gravity provides a centrifugal contribution, making the turn more difficult.  Equation \eqref{bottom_radius} has a singular point for a value of speed $v_{lim}$: below this speed, the drone is not able to steer in the lower part of trajectory, because the available centripetal force is not large enough to compensate weight. In this case though, the turning radius decreases with larger speed values.

\begin{table}
\caption{Parameters used in for steering authority analysis}
\label{Tab:radius_analysis_parameters}
\centering
\begin{tabular}{lcc}
\textbf{Parameter} & \textbf{Unit} & \textbf{Value}\\
\toprule
wing surface 		& $m^2$		&	0.21 \\
lift coefficient 	& $-$		&	1 \\
drone mass			& $Kg$		&	11 \\
\bottomrule
\end{tabular}
\end{table}

\subsection{Conclusions}
\begin{figure}
\centering
\includegraphics[width=1\columnwidth, keepaspectratio]{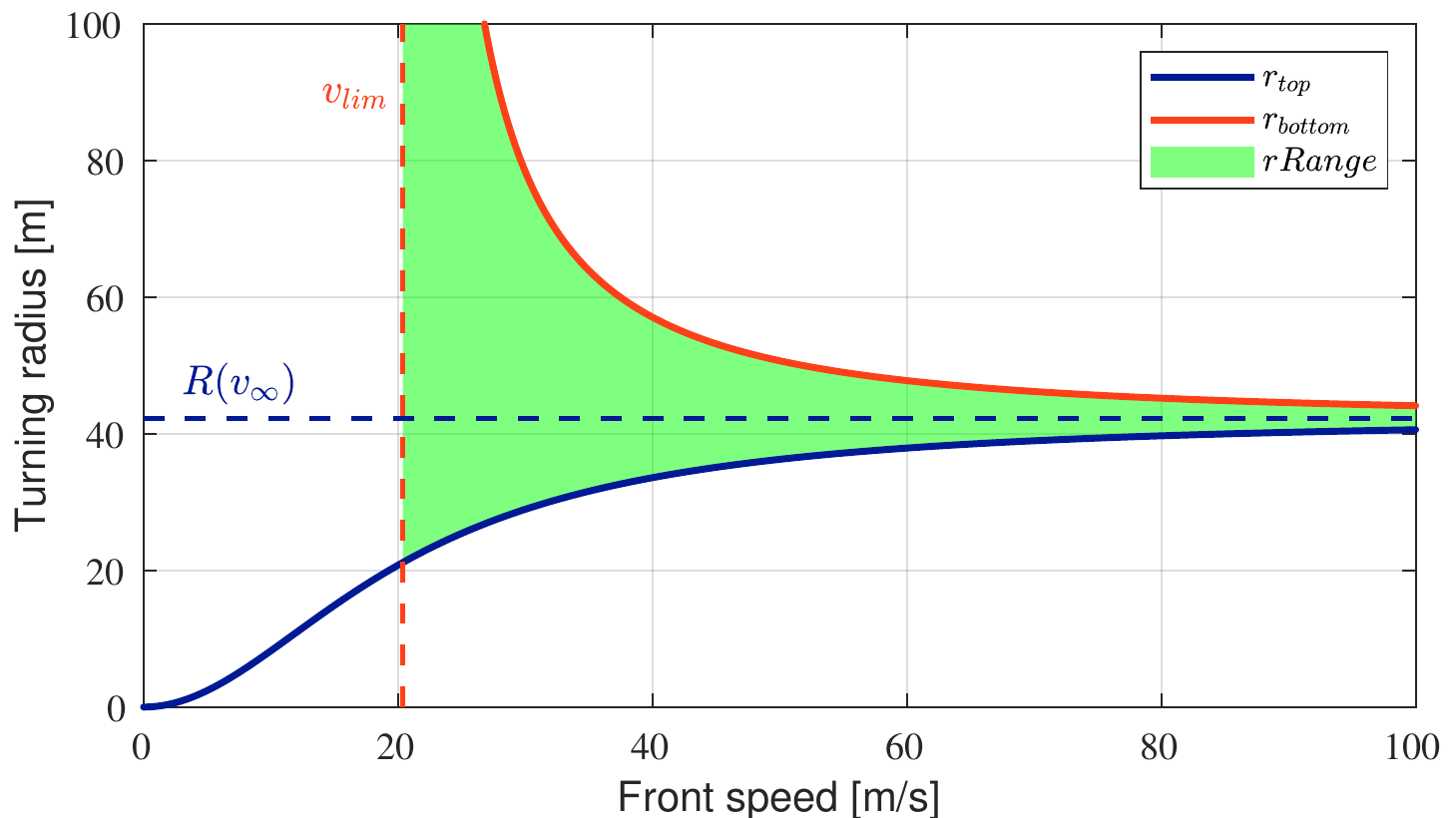}
\caption{Example of relation between turning radius and front speed for a path perpendicular to ground, for the Skypull system considered in this study. In blue, the turning radius for the upper part of the trajectory, in red for the bottom part. The green area represents the values of the turning radius during the whole maneuver, for each front speed value. Drone parameters reported in Table \ref{Tab:radius_analysis_parameters}.}
\label{SKP125_radius}
\end{figure}

From the previous analysis, the following considerations emerge:
\begin{itemize}
\item  During trajectories perpendicular to ground, the turning radius is bounded between extreme values: $r_{top}<r<r_{bottom}$;
\item For trajectories parallel to ground, the speed of the drone does not affect the steering authority. Instead, for trajectories perpendicular to ground, the steering authority is highly affected by the drone speed and it approaches a unique asymptotic value with larger speed, either from above (in the top part of the trajectory) or from below (in the bottom part). This effect is presented in Figure \ref{SKP125_radius}.
\item In a drone with lower mass, higher lift, or larger effective surface, the singularity point $v_{lim}$ is shifted to lower values, with an increase of the steering authority.
\end{itemize}

\begin{figure}
\centering
\includegraphics[width=1\columnwidth, keepaspectratio]{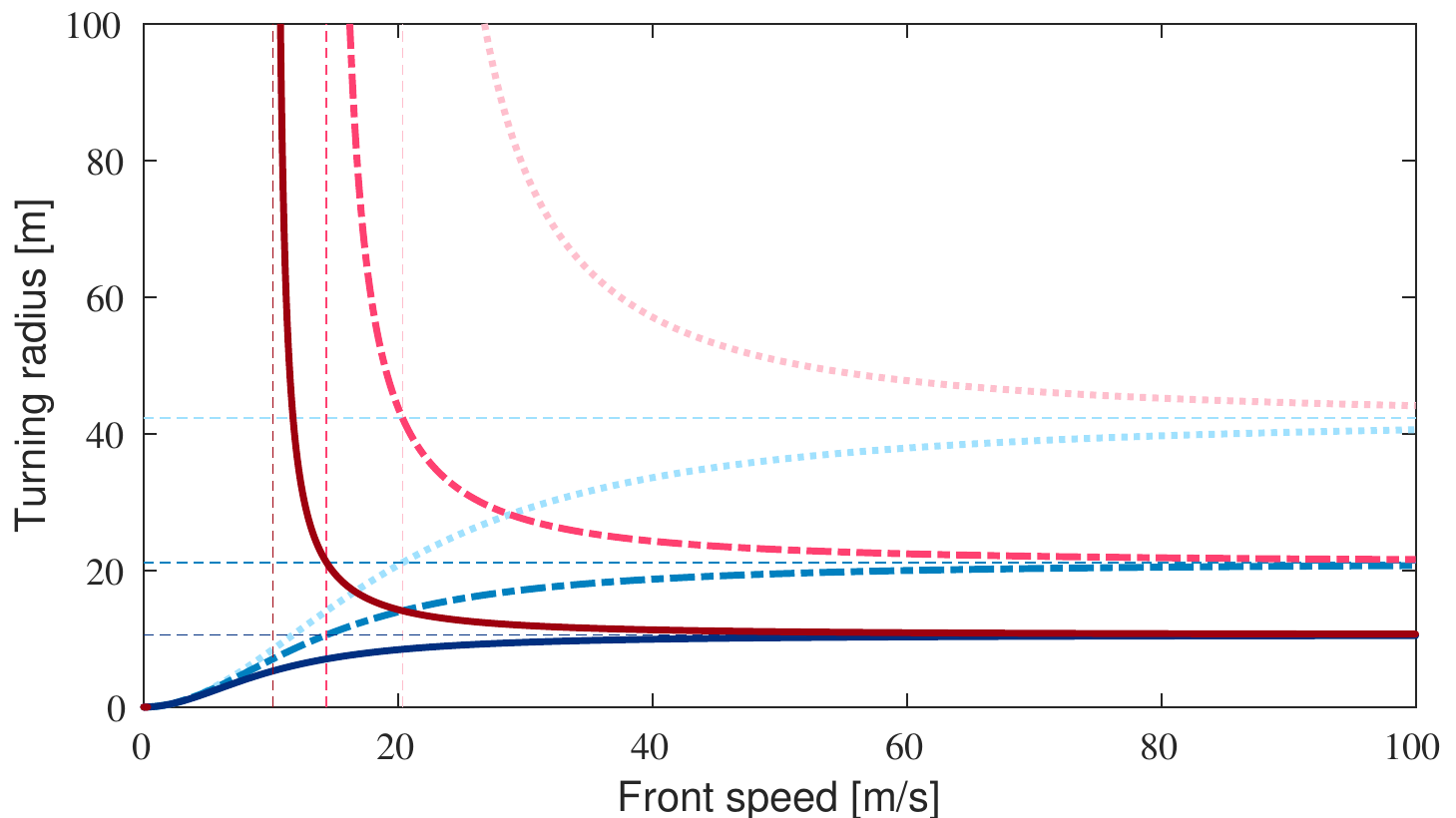}
\caption{Turning radii of the Skypull system considered in this study as a function of front speed, for the baseline total lateral surface (dotted line), and for increased surface values: twice the baseline (dash-dotted), and four times the baseline (solid). Drone parameters reported in Table \ref{Tab:radius_analysis_parameters}.}
\label{Radius_sensitivity}
\end{figure}

Figure \ref{Radius_sensitivity} shows the sensitivity of the turning radius with respect to the effective lateral wings' surface, keeping constant all other parameters. A larger surface decreases the turning radius in all conditions and reduces the minimum front speed required to steer at bottom of the path. From this analysis, it can be noted that doubling the lateral surfaces has a large impact on the steering authority, while further doubling it has a relatively lower effect, since the relationship is hyperbolic (see, e.g., \eqref{yaw_radius}). Moreover, this simplified analysis is  not valid anymore for small radius values (as compared with the drone's wingspan), since it does not take into account that when the turn is too narrow the lift distribution along the wing becomes highly uneven, and the total lift decreases significantly.\\
Based on the presented analysis, in our research we eventually decided to adopt a yaw-based strategy and to double the lateral wing surfaces with respect to the original drone design, in order to improve maneuverability.

\section{System model}\label{S:model}
In this section, we describe the models of the drone and of the ground station that we use to simulate the system. For control design, described in Section \ref{S:control}, we used different simplified models, hence introducing a mismatch between the latter and the plant model used to evaluate the control system behavior.

\subsection{Reference frames}\label{SS:frames}
\begin{figure}
	\centering
	\includegraphics[width=\columnwidth, keepaspectratio]{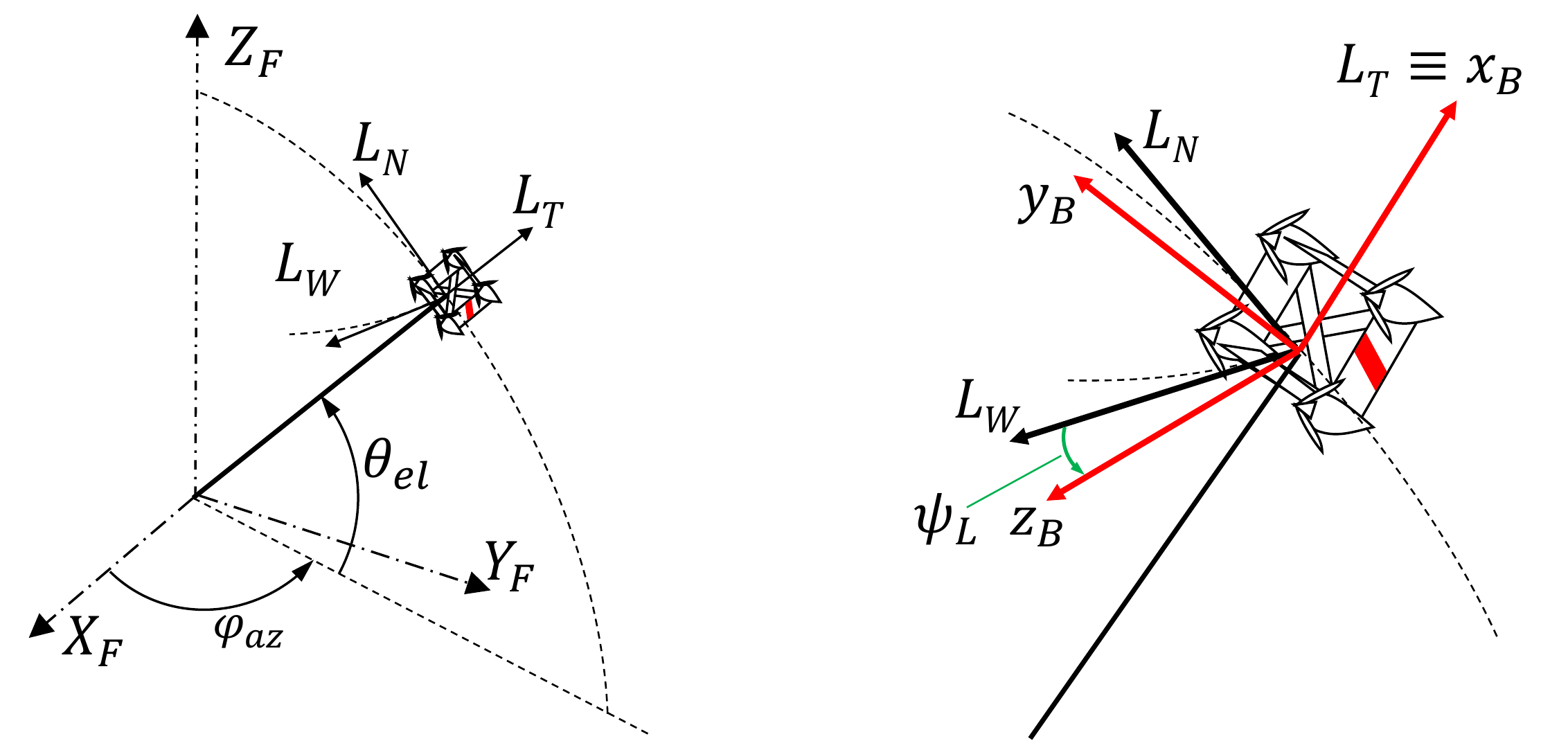}
	\caption{Left: spherical coordinates. The drone position can be expressed in terms of distance from ground station, elevation angle $\theta_{el}$, and azimuth angle $\varphi_{az}$. A sketch of the local  reference system $L$ is presented as well: $L_t$ is directed along the tether, $L_n$ points to the local North (zenith) and $L_w$ points the local West. Right: close-up view of the local reference system, and Euler angle $\psi_{L}$.}
	\label{spherical}
\end{figure}

The considered model is derived from established 6-dof aircraft model equations \cite{etkin}, with  modifications due to the box-wing architecture of the drone and the presence of the four propellers. In the remainder, $\cdot^T$ denotes the matrix transpose operation.\\
We start by introducing the drone's position $\vec{p}_{(F)}=[x_F,y_F,z_F]^T$ in the inertial reference frame $F$, which is fixed to a point on ground, with component $z_F$ pointing up and $x_F$ along a chosen direction (usually the prevalent wind direction at the considered location). For taut-tether control purposes, it is useful to convert the coordinates of the inertial system into  spherical ones:
\begin{equation}\label{E:spherical}
\begin{bmatrix} d \\ \theta_{el} \\ \varphi_{az} \end{bmatrix}=
\begin{bmatrix}
\sqrt{x_F^2+y_F^2+z_F^2} \\
atan(\frac{z_F}{\sqrt{x_F^2+y_F^2}}) \\
atan2(y_F,x_F)
\end{bmatrix}
\end{equation}
where $d$ is the drone distance from ground station, $\theta_{el}$ is the elevation angle and $\varphi_{az}$ is the azimuth angle (Figure \ref{spherical}, left).\\
For convenience when describing the control approach, we introduce three additional right-handed reference frames:
\begin{itemize}
\item $B$ $(x_B, y_B, z_B)$  is the body reference frame, fixed to the UAV and with origin in its center of gravity (Figure \ref{F:reference}). The $z_B$ axis is aligned with propellers' axes, while $x_{B}$ points towards the upper wing. This reference system is aligned with the inertial one, when the UAV is stationary in hovering mode.
\item $A_w$ $(x_{A_w}, y_{A_w}, z_{A_w})$  is the apparent wind reference system, with $x_{A_w}$ aligned with the apparent wind direction, $y_{A_w}$ parallel to the upper wing, and $z_{A_w}$ pointing up.
\item $L$ $(L_T,L_N,L_W)$ is the local (or tether) reference frame, fixed to the UAV and with $L_T$ aligned with tether, pointing opposite to the ground station, $L_N$ is pointing to the local North while $L_W$ is pointing to local West (see Figure \ref{spherical}, left).
\end{itemize}

The relative orientation between frames $F$ and $B$ can be expressed by a quaternion $\vec{q}(t)=\begin{bmatrix} q_{1}(t) & q_{2}(t) & q_{3}(t) & q_{4}(t) \end{bmatrix}^T$, where $t\in\mathbb{R}$ is the continuous time variable. For the sake of notational simplicity, in the remainder we omit the explicit dependency of the various quantities on $t$ when this is clear from the context, and we point out the constant (i.e. not time-dependent) quantities and parameters.\\
A vector in the $F$ system can be expressed in the $B$ system by means of the following rotation matrix:
\begin{equation}\label{rotation}
H_{BF}=2
\begin{bmatrix}
(q_{1}^2+q_{2}^2)-1 & (q_{2}q_{3}+q_{1}q_{4}) & (q_{2}q_{4}-q_{1}q_{3}) \\
(q_{2}q_{3}-q_{1}q_{4}) & (q_{1}^2+q_{3}^2)-1 & (q_{3}q_{4}+q_{1}q_{2}) \\
(q_{2}q_{4}+q_{1}q_{3}) & (q_{3}q_{4}-q_{1}q_{2}) & (q_{1}^2+q_{4}^2)-1]
\end{bmatrix},
\end{equation}
and a vector expressed in $A_w$ reference can be expressed in $B$ system by the following rotation matrix \cite{etkin}:
\begin{equation}
H_{WB}=
\begin{bmatrix}
\cos{\alpha} & 0 & -\sin{\alpha} \\
\sin{\alpha}\sin{\beta} & \cos{\beta} & \cos{\alpha}\sin{\beta} \\
\sin{\alpha}\cos{\beta} & -\sin{\beta} & \cos{\alpha}\cos{\beta}
\end{bmatrix}
\end{equation}
where, as already introduced in Section \ref{S:steering}, $\alpha$ is the drone's angle of attack and $\beta$ is the side-slip angle. Finally, the rotation matrix from $F$ to $L$ reads:
\begin{equation}
H_{FL}=
\begin{bmatrix}
-\sin{\varphi_{az}}\sin{\theta_{el}} 	& -\cos{\theta_{el}} 		& 	\cos{\varphi_{az}}\sin{\theta_{el}} \\
\cos{\varphi_{az}}					& 	0					&	\sin{\varphi_{az}} \\
\sin{\varphi_{az}}\cos{\theta_{el}} 	&  \sin{\theta_{el}} 		& 	\cos{\varphi_{az}}\cos{\theta_{el}}
\end{bmatrix}
\end{equation}
\subsection{Drone model}\label{SS:drone_model}
\begin{figure}
\centering
\includegraphics[width=.8\columnwidth, keepaspectratio]{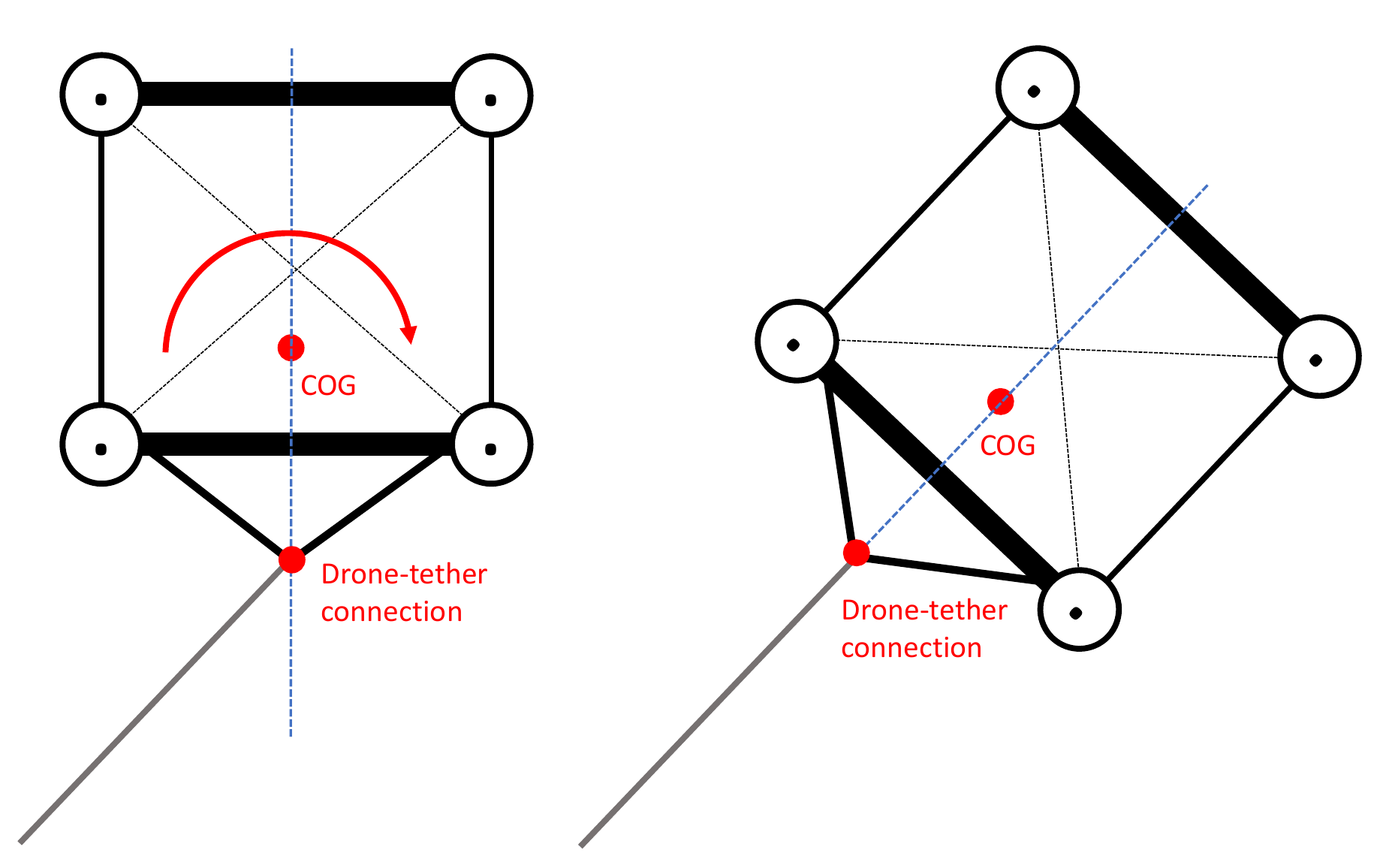}
\caption{Sketch of the tether connection. The pulling force generates both forces and moments on the drone's center of gravity.}
\label{F:tether_connection}
\end{figure}

\begin{figure}
	\centering
	\includegraphics[width=.8\columnwidth, keepaspectratio]{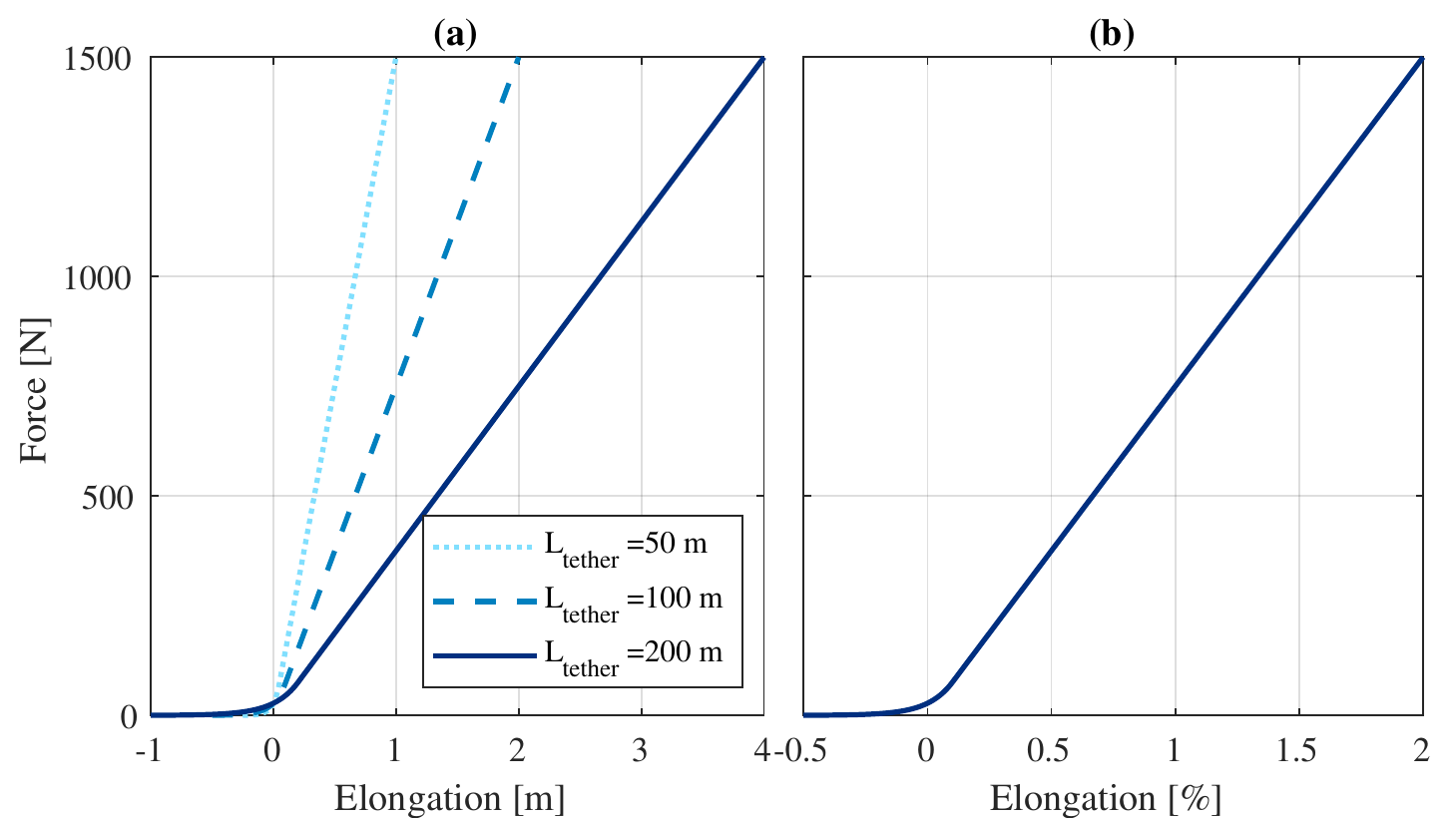}
	\caption{Tether force as a function of (a) absolute elongation  and (b) relative elongation }
	\label{F:Tether_e_k}
\end{figure}

Due to the  unbalance in mass distribution mentioned in Section \ref{S:sys_descr}, the constant inertia matrix computed with respect to the $B$ frame has non-zero terms out of the diagonal: 
\begin{equation}
I=
\begin{bmatrix}
I_{xx}&0&I_{zx} \\
0 & I_{yy} & 0 \\
I_{zx} & 0 & I_{zz}
\end{bmatrix} 
\end{equation}
The external forces and moments acting on the UAV considered in the model are:
\begin{itemize}
	\item Gravitational force;
	\item Propellers' forces and moments;
	\item Aerodynamic force and moments;
	\item Tether force and moments.
\end{itemize}
The gravitational force is computed in the inertial frame as $\vec{F}_{g (F)}=m \begin{bmatrix}0 & 0 & g\end{bmatrix}^T$, where the constants $m$ and $g$ are, respectively, the drone's mass and the gravity acceleration.\\
Each propeller  generates a thrust force, $\vec{F}_{p (B)j}$, and a drag torque in  $z_B$ direction, $T_{p (B)j},\,j=1\ldots,4$, expressed in body frame as:
\begin{equation} 
\begin{array}{cc}
\vec{F}_{p (B)j}= \begin{bmatrix}0 & 0 & b_{j}\omega_{j}^2\end{bmatrix}^T \\ 
\vec{T}_{p (B)j}= \begin{bmatrix}0 & 0 & c_{j}\omega_{j}^2\end{bmatrix}^T
\end{array} 
\end{equation}
Where $\omega_{j}$ is the rotational speed the $j^{th}$ propeller, and $b_{j},\,c_{j}$ are constant parameters.
The propellers' forces can be linearly combined to obtain the total thrust in $z_{B}$ direction, denoted by $U_{1}$, and rotational moments around $x_{B}$ ($\Delta U_{p2}$), $y_{B}$ ($\Delta U_{p3}$) and $z_{B}$ ($\Delta U_{p4}$):
\begin{equation}\label{mixer}
\begin{bmatrix}
U_{1} \\ \Delta U_{p2} \\ \Delta U_{p3} \\ \Delta U_{p4}
\end{bmatrix}
= \begin{bmatrix}
b_{1} & b_{2} & b_{3} & b_{4} \\
b_{1}d_{x_B1} & -b_{2}d_{x_B2} & -b_{3}d_{x_B3} & b_{4}d_{x_B4} \\
b_{1}d_{y_B1} & -b_{2}d_{y_B1} & b_{3}d_{y_B3} & -b_{4}d_{y_B4} \\
-c_{1} & -c_{2} & c_{3} & c_{4}
\end{bmatrix}
\begin{bmatrix}
\omega_{1}^2 \\
\omega_{2}^2 \\
\omega_{3}^2 \\
\omega_{4}^2
\end{bmatrix}
\end{equation}
where $d_{x_Bi},\,d_{y_Bi},\,i=1,\ldots,4$ are the constant position coordinates of each propeller with respect to the origin of $B$.\\ 
When deriving the model of the drone's aerodynamics, we took into account the various application points of the aerodynamic force and moment vectors generated by the four wings. Moreover, the angle of attack of one wing corresponds to the side-slip angle for the perpendicular one, thus resulting in a rather complex computation of the overall forces and moments acting at each instant on the drone.  As a matter of fact, the overall aerodynamic force and moment vectors are nonlinear mappings whose inputs are the drone's angle of attack and side-slip, the apparent wind speed vector, and the actuators' position. In the apparent wind reference system $(A_w)$, the aerodynamic force vector can be expressed as:
\begin{equation}
\vec{F}_{aero (A_w)}=\begin{bmatrix} F_{D} & F_{S} & F_{L}\end{bmatrix}^T
\end{equation}
where $F_{D},\,F_{S},\,F_{L}$ are the drag component, lateral component, and lift component, respectively, computed as:
\begin{equation} \label{aero}
\begin{array}{ccc}
F_{L}= \frac{1}{2}\rho S c_{L}(\alpha,\beta)W_{a}^2 \\ 
F_{D}=\frac{1}{2}\rho S c_{D}(\alpha,\beta)W_{a}^2 \\ 
F_{S}= \frac{1}{2}\rho S c_{S}(\alpha,\beta)W_{a}^2 
\end{array} 
\end{equation}
Where $\rho$ is the air density, $S$ is the wing area (both constant parameters), $W_{a}$ is the magnitude of the apparent wind seen by the UAV, $c_{L}$, $c_{D}$ and $c_{S}$ are respectively the lift, drag and side forces' coefficients, which depend on $(\alpha,\beta)$. These lumped parameters take into account the specific geometry of the wing. Similarly, the overall aerodynamic moment is denoted as  $\vec{M}_{aero}$ and it is computed in frame $A_w$ as a quadratic function of $W_{a}$ and a nonlinear function of $(\alpha,\beta)$, featuring suitable constant aerodynamic coefficients. The apparent wind vector $\vec{W}_{a}$ can be computed in the inertial frame as:
\[
\vec{W}_a=\vec{W}_{(F)}-\left[
\begin{array}{c}
\dot{x}_F\\
\dot{y}_F\\
\dot{z}_F
\end{array}\right]
\]
where $\vec{W}_{(F)}$ is the absolute wind vector.\\
Regarding the tether, its connection point to the drone is located on a rigid frame attached below the box-wing aircraft (see Figure \ref{F:tether_connection}).  So, the pulling force also produces a turning moment on the drone. Computing forces with respect to the body reference frame we have:
\begin{equation}
\vec{F}_{tether(B)}=-\vec{d}_{tether (B)}\vert \vec{F}_{tether}\vert
\end{equation}
where $d_{B}^{tether}$ is the tether direction, computed as (assuming taut tether):
\begin{equation}
\vec{d}_{tether (B)}=H_{BF}^T \vec{p}_{(F)}.
\end{equation}
The moment generated by the tether force is:
\begin{equation}
\vec{M}_{tether(B)}=\vec{p}_{tether(B)}  \wedge \vec{F}_{tether(B)}
\end{equation}
where $\vec{p}_{tether(B)}$ is the position of the connection point in body reference frame, relative to the center of gravity of the drone.
The force magnitude $\vert \vec{F}_{tether}\vert$ is computed considering non-zero load transfer only with taut tether:
\begin{equation}\label{tether_equation}
\vert \vec{F}_{tether}\vert=\max{(0,k_{s}(L_{tether})(d-L_{tether}))}
\end{equation}
where $k_{s}$ is the tether stiffness, $d$ is the drone distance from the GS and $L_{tether}$ is the unrolled tether length, i.e. $(d-L_{tether})$ is the tether elongation. Tether stiffness is a parameter that depends on the length of unrolled tether:
\begin{equation}\label{stiffness}
k_{s}(L_{tether})=\frac{F_{max}}{\Delta L_{max}}=\frac{F_{max}}{\epsilon_{max}L_{tether}}
\end{equation}
where $F_{max}$ is the maximum force that the tether can withstand, $\Delta L_{max}$ is the corresponding  maximum elongation, and $\epsilon_{max}$ is the maximum elongation relative to the  length. Being $\frac{F_{max}}{\epsilon_{max}}$ a constant parameter for a given tether, the stiffness depends only on its length, and decreases with it according to the hyperbolic equation \eqref{stiffness}.\\
To avoid numerical solution problems, in numerical integration routines it is advisable to implement an approximated version of \eqref{tether_equation}:
\begin{equation}\label{stiffness2}
\vert \vec{F}_{tether}\vert= \begin{cases}
\frac{F_{max}}{\epsilon_{max}}\epsilon, & \mbox{if } \epsilon \ge \epsilon_0 \\
\frac{F_{max}\epsilon_{0}}{\epsilon_{max}e}e^{\frac{\epsilon}{\epsilon_0}}, & \mbox{if } \epsilon<\epsilon_0
\end{cases}
\end{equation}
This formulation is actually identical to \eqref{stiffness} for positive elongation, and uses a decreasing exponential approximation for negative elongation. This allows one to avoid the non-differentiable point at zero elongation, which is instead present in \eqref{tether_equation}. The resulting curve is shown in Figure \ref{F:Tether_e_k}.
Finally, tether drag is accounted for as an incremental term that increases  the drone aerodynamic drag:
\begin{equation}\label{eq:drag_total}
c_{D_{tot}}=c_D+\frac{c_{D,l} d_t L_t}{8 S_{up}}
\end{equation}
where $c_D$  is the drone drag coefficient, $c_{D,l}$ is the tether drag coefficient, $S_{up}$ is the total aerodynamic upper surface, $d_t$ and $L_t$ are respectively the tether diameter and the unrolled length. The equation is derived from a momentum balance considering a linear dependency of the apparent speed seen by each infinitesimal segment of tether, and the distance of such a segment from the ground station, see e.g. \cite{Fagiano2009}.\\

\noindent Considering all of the described external forces and moments, the resulting drone model consists of 13 ordinary differential equations (ODEs):
\begin{equation}\label{dynamic}
\begin{array}{rcl}
F_{x_B}&=&m(\dot{U}-RV+QW) \\ 
F_{y_B}&=&m(\dot{V}-PW+UR) \\ 
F_{z_B}&=&m(\dot{W}-QU+PV) \\ 
M_{x_B}&=&\dot{P}I_{xx}-(\dot{R}+PQ)I_{zx}+RQ(I_{zz}-I_{yy})+Q \sum_{j=1}^{4} {h_{z,j}} \\
M_{y_B}&=&\dot{Q}I_{yy}+({R}^2+{P}^2)I_{zx}+PR(I_{xx}-I_{zz})-P\sum_{j=1}^{4} {h_{z,j}} \\
M_{z_B}&=&\dot{R}I_{zz}-(\dot{P}+QR)I_{zx}+PQ(I_{yy}-I_{xx}) \\
\begin{bmatrix} 
\dot{q_{1}} \\
\dot{q_{2}} \\
\dot{q_{3} }\\
\dot{q_{4}}
\end{bmatrix}
&=& \frac{1}{2}
\begin{bmatrix}
0&-P&-Q&-R \\
P & 0 & R & -Q \\
Q & -R & 0 & P \\
R & Q & -P & 0
\end{bmatrix}
\begin{bmatrix}
q_{1} \\
q_{2} \\
q_{3} \\
q_{4}
\end{bmatrix}
\\
\begin{bmatrix}
\dot{x_F} \\
\dot{y_F} \\
\dot{z }
\end{bmatrix}
&=& H_{BI}
\begin{bmatrix}
U \\
V \\
W
\end{bmatrix}
\end{array}
\end{equation}
where  $U$, $V$ and $W$ are the drone's velocity vector components in the body frame $B$; $ P$, $Q$ and $R$ are the rotational speeds around the axes of $B$; $F_{xB}$, $F_{yB}$ and $F_{zB}$ are the components in the $B$ frame of the vector sum of all external forces acting on the UAV; $M_{xb}$, $M_{yb}$ and $M_{zb}$ are the components in the $B$ frame of the vector sum of all external moments applied to the UAV. Finally, the constant parameters $h_{z,j},\,j=1,\ldots,4$ are the moments of inertia of motors/propellers in $z_B $ direction. The control inputs are the total thrust force, $U_1$, and the total moments provided by the propellers and by the aerodynamic control surfaces, denoted as $U_2,\,U_3$ and $U_4$ for rotations around axes $x_{B},\,y_B,$, and $z_B$, respectively. These variables do not appear explicitly in \eqref{dynamic}, since they contribute, respectively, to $F_{zb}$, $M_{xb}$, $M_{yb}$ and $M_{zb}$. In particular, the  turning moments $U_2,\,U_3$ and $U_4$ are given by the moments $\Delta U_{p2},\,\Delta U_{p3},\,\Delta U_{p4}$, provided by the propellers (see \eqref{mixer}) and exploited mainly in hovering mode, plus the contributions of the aerodynamic surfaces, used prevalently in dynamic flight mode. The latter are modeled here directly in terms of control moments. In practice, such moments are nonlinear functions of the positions of the discrete control surfaces that are present on each wing. The precise characterization of such moments as a function of flight conditions and surface position is a proprietary know-how of Skypull SA, not disclosed here for confidentiality reasons. On the other hand, considering directly the control moments $U_2,\,U_3$ and $U_4$ results in a more general approach that can be applied to other drone types, by inserting the input mapping between discrete control surface positions and resulting moments.

\subsection{Ground station model}\label{SS:GS_model}

From rotational equilibrium, the model of the motor-winch subsystem reads:
\begin{equation}\label{GS_equation}
J \ddot{\lambda}	=	\vert \vec{F}_{tether}\vert r_w - T - \beta_f \dot{\lambda}
\end{equation}
where $J$ is the total moment of inertia of the winch and the motor, $\lambda$ is the angular position of the winch, $r_w$ is the winch radius, $T$ is the torque applied by the electric machine to the winch and $\beta_f$ is the viscous friction coefficient.

\section{Control design}\label{S:control}
\begin{figure}
	\centering
	\includegraphics[width=.8\columnwidth, keepaspectratio]{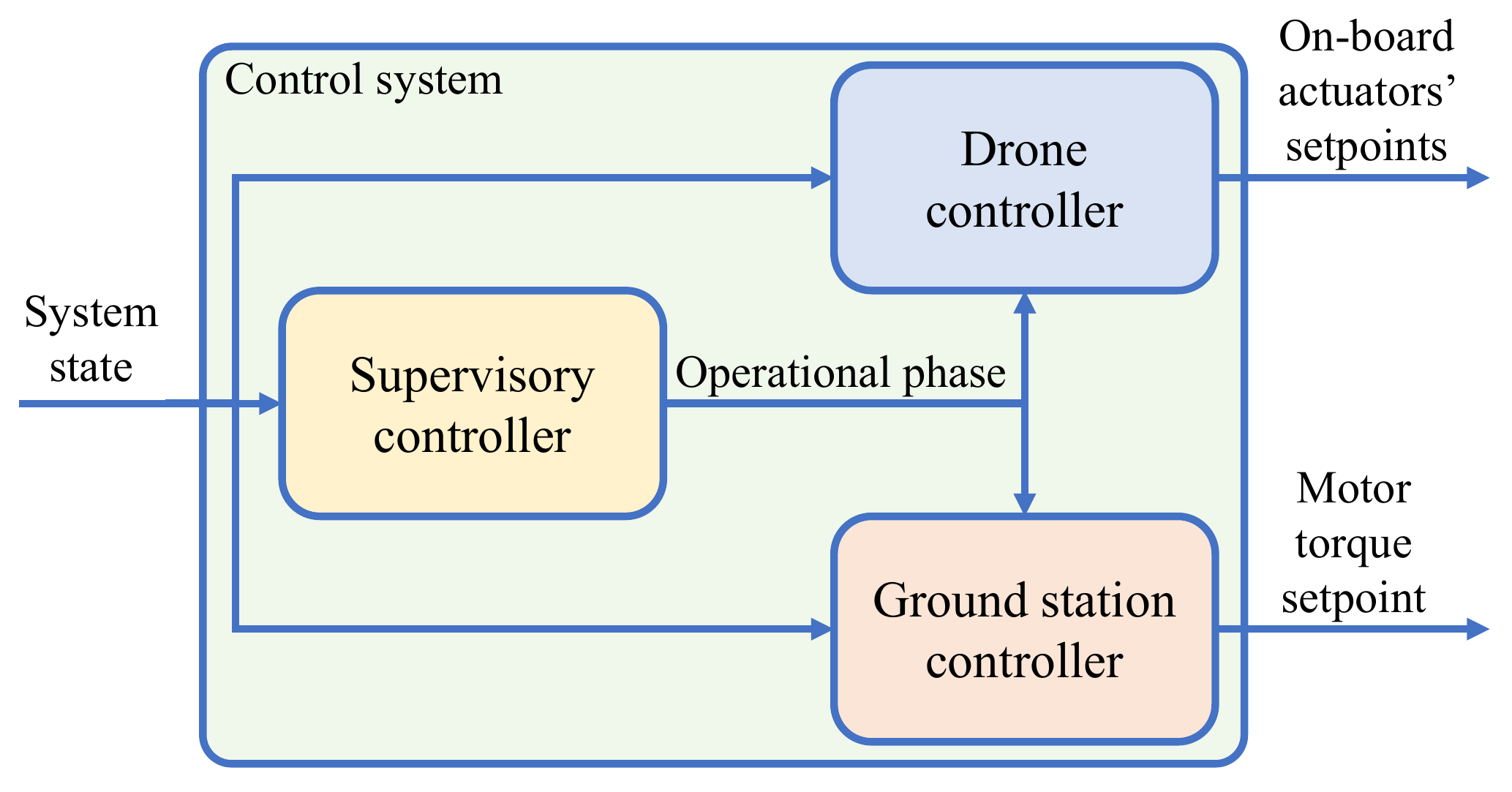}
	\caption{Layout of the control system}
	\label{F:control_topology}
\end{figure}

We employed simplified models of the system's behavior to design the various control strategies, which will be recalled in this section. The final control approach has been then tested on the full nonlinear model \eqref{E:spherical}-\eqref{GS_equation}, which captures quite accurately all the relevant effects occurring in the real system.\\ We start our description from the overall layout of the control logic, presented in Figure \ref{F:control_topology}. 
A supervisory controller is in charge of monitoring the execution of the current phase and of switching to the next one. In each phase, the supervisor issues a corresponding operational mode to the drone's and the ground station's controllers, which in turn feature a hierarchical topology. In the next sections, we describe each one of the involved control functions. We assume that the full state of the drone and of the ground station is available: this is reasonable considering that each state variable is measured with redundant sensors, and filtering algorithms are in place as well to reduce the effects of noise.

\subsection{Drone Controller}\label{SS:Drone}
For the sake of drone control design, it is useful to distinguish the following three different working conditions:
\begin{enumerate}
\item Hovering;
\item Dynamic flight with slack tether (free-flight);
\item Dynamic flight with taut tether (taut-tether flight).
\end{enumerate}
To formulate the feedback control algorithms in an intuitive way, we resorted to three definitions of Euler angles, one for each working condition, expressed as functions of the entries of the rotation matrix $H_{BF}$ \eqref{rotation}:
\begin{itemize}
	\item Euler angles for hovering $ (\varphi_{h},\theta_{h},\psi_{h})$, defined as $Z-Y-X$ right handed rotation sequence from the inertial frame $F$ to body 				frame $B$:
	\begin{equation}\label{eq:euler-ho}
	\begin{bmatrix}
	\varphi_{h}\\ \theta_{h} \\ \psi_{h}
	\end{bmatrix}
	=
	\begin{bmatrix}
	\arctan{\frac{H_{BF} (2,3)}{H_{BF} (3,3)}} \\
	\arcsin{-H_{BF} (1,3)} \\
	\arctan{\frac{H_{BF} (1,2)}{H_{BF} (1,1)}}
	\end{bmatrix}
	\end{equation}
	These are the typical angles used for multicopter control. 
	\item Euler angles for free-flight $(\varphi_{f},\theta_{f},\psi_{f})$, defined as $X-Y-Z$ right handed rotation sequence from $F$ to $B$:
	\begin{equation}\label{eq:euler-ff}
	\begin{bmatrix}
	\varphi_{f}\\ \theta_{f} \\ \psi_{f}
	\end{bmatrix}
	=
	\begin{bmatrix}
	\arctan{\frac{H_{BF} (3,2)}{H_{BF} (3,1)}} \\
	\arcsin{-H_{BF} (3,3)} \\
	\arctan{\frac{H_{BF} (2,3)}{H_{BF} (1,3)}}
	\end{bmatrix}
	\end{equation}
	With this convention, when the drone flies at zero pitch and roll, the $Z_B$ axis is parallel to the ground and $X_B$ axis points up. These angles correspond to the typical ones for airplane control.
	\item Euler angles for taut-tether flight $(\varphi_{L},\theta_{L},\psi_{L})$. These are defined similarly to the Euler angles for free-flight mode, but referring to an initial condition in which the drone $x_B$ axis is aligned with the tether, which corresponds to $\varphi_{L}=\theta_{L}=0$. In these conditions, the yaw angle $\psi_{L}$ can assume any value, depending on the direction the drone is pointing to (see Figure \ref{spherical}, left):
	\begin{equation}
	\begin{bmatrix}\label{eq:euler-tt}
	\varphi_{L}\\ \theta_{L} \\ \psi_{L}
	\end{bmatrix}
	=
	\begin{bmatrix}
	\arctan{\frac{-H_{BL} (1,2)}{H_{BL} (1,1)}} \\
	\arcsin{H_{BL} (1,3)} \\
	\arctan{\frac{-H_{BL} (2,3)}{H_{BL} (3,3)}}
	\end{bmatrix}
	\end{equation}
	where $H_{BL}=H_{FL}\cdot H_{BF}$ is the rotation matrix between body axis and local reference frame.
\end{itemize}
The use of two different sets of Euler angles for hovering and free-flight (\eqref{eq:euler-ho} and \eqref{eq:euler-ff}, respectively) also allows us to avoid the well-known gimbal lock problem, occurring at pitch angle $\theta=\frac{\pi}{2}$. Indeed, such a pitch value can occur in either one of the two Euler triplets 1. and 2., but never in both triplets at the same time. The Euler angles for tethered flight mode \eqref{eq:euler-tt} are further introduced in order to describe in an intuitive way the misalignment of the drone's $x_B$ axis with respect to the tether, hence simplifying the design of the attitude controllers.

\begin{figure}
\centering
\includegraphics[width=1\columnwidth, keepaspectratio]{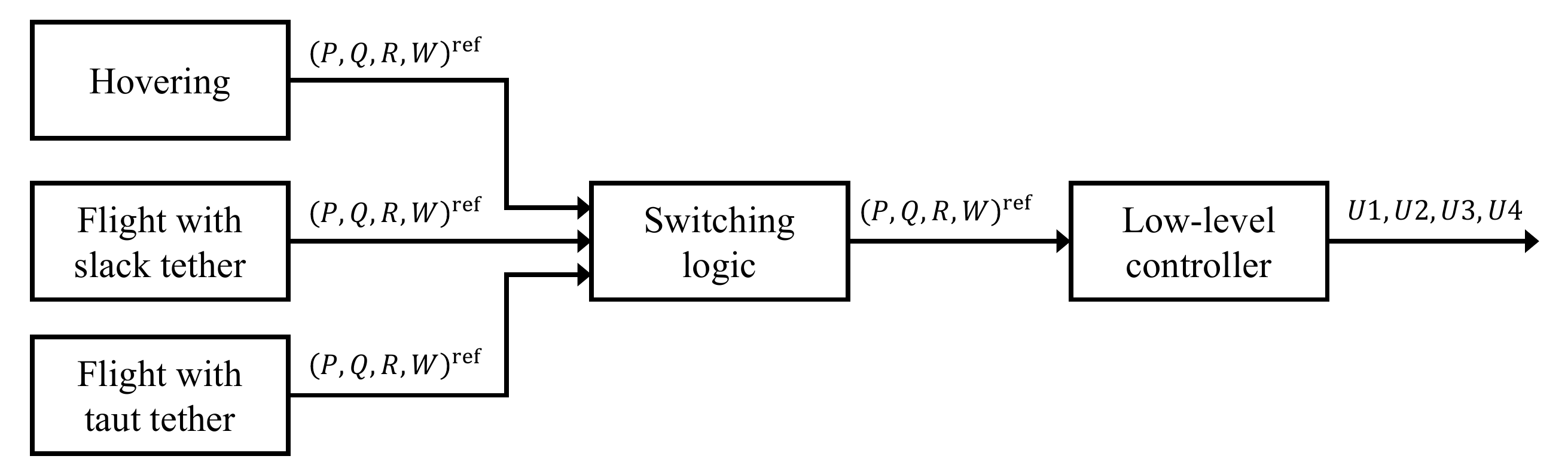}
\caption{General scheme of the drone controller.}
\label{schema}
\end{figure}
The drone controller consists of a hierarchical structure shown in Figure \ref{schema}, with a common low-level attitude controller and different high-level controllers for the various working conditions.  In Section \ref{SS:low_level} we present the low-level attitude controller and the high-level ones for hovering and free-flight, while in Section \ref{SS:tethered_control} we describe the one for taut-tether flight. In the remainder, all the gains of the feedback controllers are computed via pole-placement, unless otherwise noted.
\subsubsection{Low level attitude controller, and hovering and free-flight controllers}\label{SS:low_level}
The low level controller is the same developed in \cite{Todeschini2019}. It is based on simplified equations of the rotational motion and take into account the non-symmetrical inertia matrix. \\
The hovering controller computes the reference rotation rates sent to the attitude controller when the drone is in hovering condition. This controller is based again on a hierarchical approach: the inner loop tracks the reference attitude and altitude, a mid level controller tracks the reference speed in $x_F$ and $y_F$, and an outer loop tracks the position of the drone in the $(x_F,y_F)$ plane. \\
The controller for free-flight features a hierarchical logic, with an inner loop responsible for attitude tracking, a middle loop responsible for altitude tracking, and an outer loop for inertial navigation planning.
This controller is designed using the Euler angles for dynamic flight, $(\varphi_{f},\theta_{f},\psi_{f})$. Due to the system layout and intended use of the UAV in an airborne wind energy system, the steering mechanism is a yawing motion instead of a roll one, commonly used in traditional airplanes.
For further information to these controllers we refer the reader to \cite{Todeschini2019}.

\subsubsection{Taut-tether flight controller}\label{SS:tethered_control}
This controller exploits the Euler angles for taut-tether flight, which define the rotation from the body reference frame $B$ to the local reference frame $L$. The main difference between taut-tether control and the free-flight is that we now provide switching target points in spherical coordinates, following an established approach to obtain figure of eights, see e.g. \cite{FZMK14,Erhard2015}. Assuming that the tether has instantaneously constant length, the drone is restricted to move along a spherical surface. As a consequence, any motion direction is identified by a single planar angle, called velocity angle, see e.g. \cite{FZMK14}. This is defined as:
\begin{equation}
\nu=\text{atan2}\left(\vec{\dot{p}}_{(L)}^{\,T}\,L_W,\;\vec{\dot{p}}_{(L)}^{\,T}\,L_N\right).
\end{equation}
where $\text{atan2}(\cdot,\cdot)$ is the four-quadrant arctangent function. The reference velocity angle $\nu^{(ref)}$ is  computed in order to make the drone move towards the current target point:
\begin{equation}
\nu^{(ref)}=\text{atan2}\left(-(\varphi_{az}^{(ref)}-\varphi_{az})\cos{\theta_{el}},-(\theta_{el}^{(ref)}-\theta_{el})\right)
\end{equation}
where $\theta_{el}^{(ref)},\,\varphi_{az}^{(ref)}$ are, respectively, the elevation and azimuth angles of the target point. Then, we design a feedback controller to maintain $\theta_{L}$ and $\varphi_{L}$ angles equal to zero, so to keep a stable tethered flight, and to regulate the yaw to make the drone track the reference velocity angle $\nu^{(ref)}$. Specifically, the controller takes the form:
\begin{equation} \label{Tethered_eq}
\begin{bmatrix}
\dot{\varphi_{L}} \\ \dot{\theta_{L}} \\ \dot{\psi_{L}}
\end{bmatrix}
=
\begin{bmatrix}
k_{\varphi_L} & 0 & 0 \\
0 & k_{\theta_L} & 0 \\
0 & 0 & k_{\psi_L}
\end{bmatrix}
\begin{bmatrix}
0-\varphi_L \\
0-\theta_L \\
\nu^{(ref)}-\nu
\end{bmatrix}
\end{equation}
Note that velocity angle is used instead of the true heading angle $\psi_L$ to track speed direction more precisely, as it compensates the side slip angle $\beta$.\\
Finally, the reference angular velocities around each axis, provided as reference signals to the low-level attitude controller (see Figure \ref{schema}) are computed by means of the following rotation matrix:
\begin{equation} \label{rot_tethered}
\begin{bmatrix}
P^{(ref)} \\ Q^{(ref)} \\ R^{(ref)}
\end{bmatrix}
=
\begin{bmatrix}
0 & \sin{\varphi_L} & \cos{\varphi_L}\cos{\theta_L} \\
0 & \cos{\varphi_L} & -\sin{\varphi_L}\cos{\theta_L} \\
1 & 0        &  \sin{\theta_L}
\end{bmatrix}
\begin{bmatrix}
\dot{\varphi_L} \\ \dot{\theta_L} \\ \dot{\psi_L}
\end{bmatrix}
\end{equation}

\subsection{Ground station control}\label{SS:GS_control}
The ground station must control the tether release speed and/or the applied force by means of the electric motor. We distinguish three possible working conditions:
\begin{enumerate}
\item Low  tether force;
\item Generation phase: pulling force applied on the tether and positive release speed;
\item Reentry phase with taut tether: pulling force applied on the tether and negative release speed.
\end{enumerate}
We developed three controllers in order to operate in these different working conditions.
\subsubsection{Low  tether force}
The controller in low force conditions requires an accurate position measurement of the drone in order to guarantee that the tether length is larger than the distance between the drone and the ground station, however without exceeding too much, to avoid possible tether entanglement on ground. The idea is to keep an extra-length of the released tether with respect to the drone-ground station distance, by means of a constant offset $\widehat{\Delta L}$.
For this kind of control, we used a PD controller tuned on the systems parameters, whose transfer function reads:
\begin{equation}
T_{NoTension}(s)=(K_p+\frac{K_d s}{1+\frac{K_d}{N_d}s})(L^{(ref)}-L)
\end{equation}
where $s$ is the Laplace variable, and $K_p$, $K_d$ and $N_d$ are tuning parameters based on the model equation \eqref{GS_equation}. $L^{(ref)}$ is computed by means of drone distance $|\vec{p}_{F}|$ and tuning parameter $\widehat{\Delta L}$:
\begin{equation}
L^{(ref)}=|\vec{p}_{F}|+\widehat{\Delta L}.
\end{equation}
$\widehat{\Delta L}$ is usually set at a few meters, depending on the precision of the GPS measure. To avoid tether entanglement, this controller is active only when the tether has to be reeled-in, i.e., during free-flight reentry phases and landing phases.
\subsubsection{Generation phase}
During the generation phase, the ground station controller must pull the tether in order to maximize the generated power. To this end, it can be demonstrated (see, e.g., \cite{Zgraggen2016}) that the best controller (considering only the traction phase) takes the form:
\begin{equation}
T_{generation}=k_{generation} \dot{\lambda}^2
\end{equation}
 with
\begin{equation}
k_{generation}=2 \rho S C_L r_W^2 E
\end{equation}
where $\rho$ is the air density, $S$, $C_L$ and $E$ are respectively the total aerodynamic surface the lift coefficient and the aerodynamic efficiency of the drone and $r_w$ is the winch radius.
\subsubsection{Reentry phase with taut tether}
During the retraction phase with taut tether, the ground station controller shall guarantee that the tether is reeled-in under large enough pulling force. The tether force-speed combination has an impact on the overall efficiency of the pumping cycle. The most practical solution is to employ a  speed controller with a reference reeling speed that has to be tuned to obtain a good cycle efficiency (see Section \ref{SS:cycle}).
Such a  controller is:
\begin{equation}
T_{reentry}= -k_{reentry}(\dot{L}^{(ref)}-\dot{L})
\end{equation}
where $\dot{L}^{(ref)}$ is usually set at few meters per second (at negative value).
\subsection{Supervisory controller and transition management }\label{SS:Coordination}
\begin{figure}
\centering
\includegraphics[width=\columnwidth, keepaspectratio]{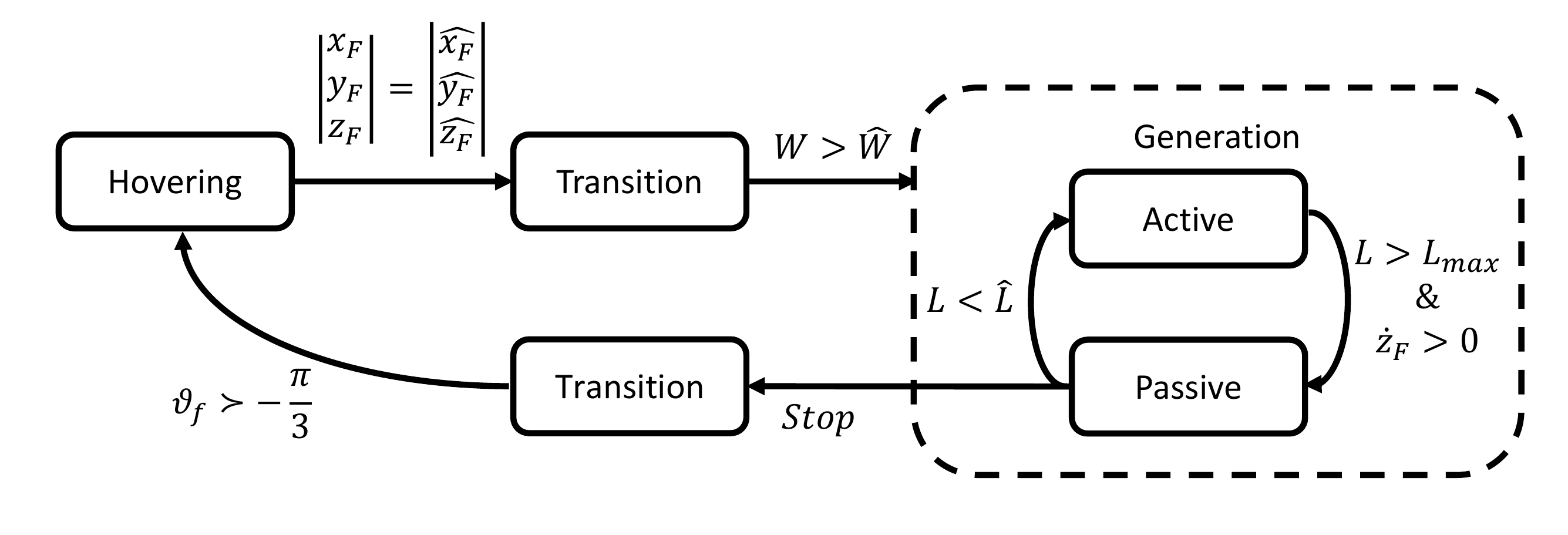}
\caption{Supervisory control strategy with switching conditions among phases.}
\label{high_level_startegy}
\end{figure}

Coordination between ground station and drone controllers is crucial to achieve an efficient and safe operation of the system. The supervisory logic depicted in Figure \ref{F:control_topology}  is responsible for the selection of the mid-level controllers used during the different phases.
According to the system's operating principle, we identified five operational phases, see Figure \ref{high_level_startegy}:
\begin{enumerate}
\item Hovering to the point in which  power generation starts (for take-off) and back to the ground station (for landing). In this condition, the hovering controller is active on the drone, and the ground station is controlled in order to have no tension on the tether. The hovering controller is commanded to move to a designated downwind position to start pumping operation;
\item Transition from hovering to taut-tether flight. In this phase, which occurs between take-off and power generation, a chosen  reference speed and reference yaw rate are provided to the hovering controller on the drone in order to make it steer into crosswind flight. Regarding the ground station, in this phase the power generation controller is enabled;
\item Tethered flight in crosswind. When the drone speed is sufficiently high, the controller for taut-tether flight is enabled on the drone, which starts to track reference points in order to accomplish figure-of-eight path. The ground station control is set to power generation;
\item Reentry phase. When the tether length exceeds a maximum value, the reentry phase starts. If the selected strategy is the reentry with taut-tether, the selected controllers for the drone and ground station are, respectively, the taut-tether flight controller and the reentry controller with taut tether. Instead, if free-flight reentry is selected, the employed controllers are the free-flight one for the drone, and the low force one for the ground station;
\item Transition from generation phase to hovering. In this phase, which occurs between power generation and landing, the ground station controller is set to low force, while the low-level attitude controller on the drone is provided with reference angles and altitude that make the aircraft pitch up, slow down and reach a hovering configuration. The hovering controller is  engaged when the system's state is close to such a configuration.
\end{enumerate}

\begin{table*}[!htb]
	\small
\centering
\begin{tabular}{|l|c|c|c|c|c|}
\cline{2-6}
\multicolumn{1}{c|}{} & \multicolumn{5}{|c|}{\textbf{Phase}}\\\cline{2-6}
\multicolumn{1}{c|}{} & \multicolumn{1}{|c|}{Hovering} & \multicolumn{1}{|c|}{Transition} & \multicolumn{1}{|c|}{Pumping} & \multicolumn{1}{|c|}{Pumping} & \multicolumn{1}{|c|}{Transition}\\
\multicolumn{1}{c|}{} & \multicolumn{1}{|c|}{} & \multicolumn{1}{|c|}{to generation} & \multicolumn{1}{|c|}{(traction)} & \multicolumn{1}{|c|}{(reentry)} & \multicolumn{1}{|c|}{to hovering}\\
\hline
\hline
\textbf{Drone}	& \multicolumn{1}{|c|}{HO} & \multicolumn{1}{|c|}{HO} & \multicolumn{1}{|c|}{TT} & \multicolumn{1}{|c|}{TT or FF} & \multicolumn{1}{|c|}{FF} \\
\hline
\textbf{Ground s.}		& \multicolumn{1}{|c|}{LF} & \multicolumn{1}{|c|}{GE} & \multicolumn{1}{|c|}{GE} & \multicolumn{1}{|c|}{TT or LF}& \multicolumn{1}{|c|}{LF}\\
\hline
\end{tabular}
\caption{Active drone and ground station controllers in the different operational phases. Drone controllers: HO=Hovering; FF=Free-flight; TT=Taut tether. Ground station controllers: LF=Low force; GE=Power generation; TT=Taut tether reentry. }
\label{active_controllers}
\normalsize
\end{table*}

Figure \ref{high_level_startegy} also shows the switching conditions between the various phases, while Table \ref{active_controllers} summarizes the drone and ground station controllers employed in each phase.
The transition from hovering to taut-tether flight is necessary to assure that the drone reaches a given speed value before starting the figure of eight tracking. This is ensured by providing a high reference front speed and enabling the tethered controller only above a chosen, large-enough speed. Instead, to have a good transition from taut-tether flight to free-flight and vice-versa, there is no need of an additional controller, because the drone has a speed of the same order of magnitude in both situations. However, an accurate tuning of the switching between the two working conditions has to be done  to ensure a smooth  transition.


\section{Simulations results}\label{S:results}
\begin{figure}
\centering
\includegraphics[width=1\columnwidth, keepaspectratio]{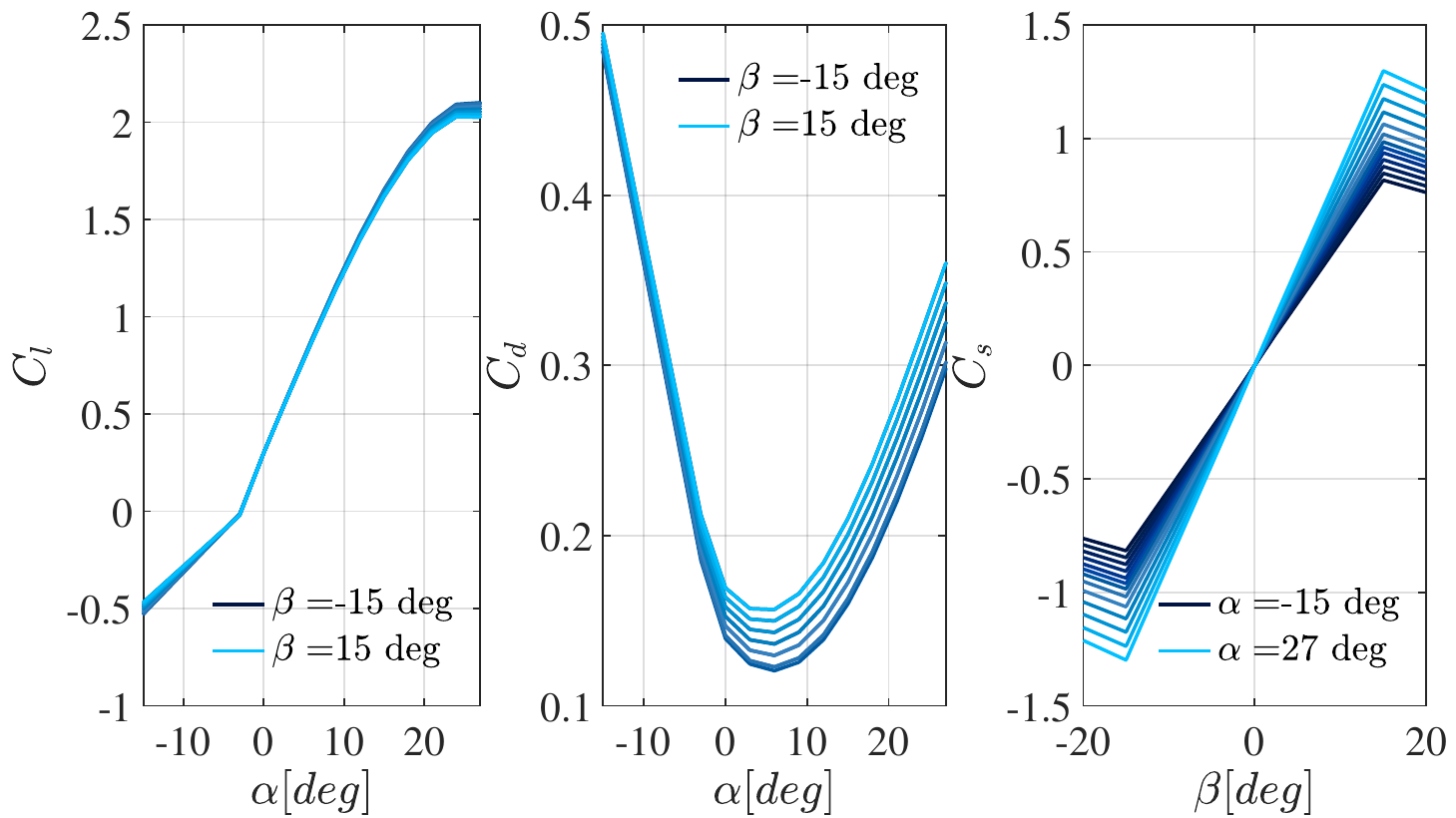}
\caption{Aerodynamic coefficients used in the simulations. lift and drag are plotted as function of angle of attack and for different side slip angle, while side coefficient is plotted function of side slip angle and for different angle of attack.}
\label{F:aero_coeff}
\end{figure}
\begin{table}
\caption{Controllers' parameters used in simulations}
\label{controllers_poles}
\centering
\begin{tabular}{|l|l|c|}
\hline
\multicolumn{2}{|c|}{$\,$} & \textbf{Position of closed loop poles}\\
 \multicolumn{2}{|c|}{\textbf{Controller}} & \textbf{(continuous time, absolute value)}\\
\hline
\multirow{3}*{Hovering} 	& Position 	& 0.2 \\
						& Speed		& 1 \\
						& Attitude	& 5 	\\
\hline
\multirow{2}*{Flight} 	& Attitude	& LQR \\
						& Altitude	& 10 \\			
\hline
Tethered		& Attitude	& 5 \\
\hline
\multirow{2}*{Low level} 	& Rotations 	& 40 \\
						& Thrust		& 20 \\
\hline
\end{tabular}
\end{table}

\begin{table*}[!htb]
\caption{Parameters used in simulations}
\label{Tab:Parameters}
\centering
\begin{tabular}{l|l|c|c}
 & \textbf{Parameter} & \textbf{Unit} & \textbf{Value}\\
\hline
\multirow{6}{*}{Drone} 			& Mass 							& $kg$				& 11.3 \\
								& Inertia matrix				& $kg \cdot m^2$ 	& $\begin{bmatrix}
                                                                                   4.53 & 0 & -1.35 \\
                                                                                    0 & 3.28 & 0 \\
                                                                                   -1.35 & 0 & 5.49
                                                                                    \end{bmatrix} $\\
								& Aerodynamic chord 			& $m$ 				& 0.18 \\
								& Wing span 					& $m$ 				& 1.17\\
								& Wing surface 					& $m^2$ 			& 0.21\\
								& Tether connection point 		& $m$ 				& $\begin{bmatrix}
				                                                    					-1 & 0 & 0
						                                            				\end{bmatrix}^T$\\
\hline
\multirow{4}{*}{Propellers} 	& Max speed						& $rpm$				& 12000\\
								& Thrust coefficient			& $\frac{N}{rpm^2}$	& 1.1e-6\\
								& Drag coefficient 				& $\frac{Nm}{rpm^2}$& 2.04e-8\\
                        		& Distances from center of mass & $m$ 				& $\begin{bmatrix}
                                                                                   0.632    & 0.395 \\
                                                                                   0.632    & 0.585 \\
                                                                                   0.632    & 0.395 \\
                                                                                   0.632    & 0.585
                                                                                    \end{bmatrix} $\\
\hline
\multirow{5}{*}{Tether} 		& Length						& $m$				& 500\\
								&	Diameter 					& $mm$				&	0.83 \\
								& Max elongation				& $\%$ 				& 1\\
					    		& Drag coefficient 				& - 				& 1\\
								& Initial length 				& $m$ 				& 5\\
\hline
\multirow{3}{*}{Winch} 			& Radius						& $m$				& 0.159\\
								& Moment of inertia				& $kg \cdot m^2$	& 0.2\\
								& Friction coefficient 			& $Nms$				& 0.01\\
\end{tabular}
\end{table*}

\begin{figure}
\centering
\includegraphics[width=1\columnwidth, keepaspectratio]{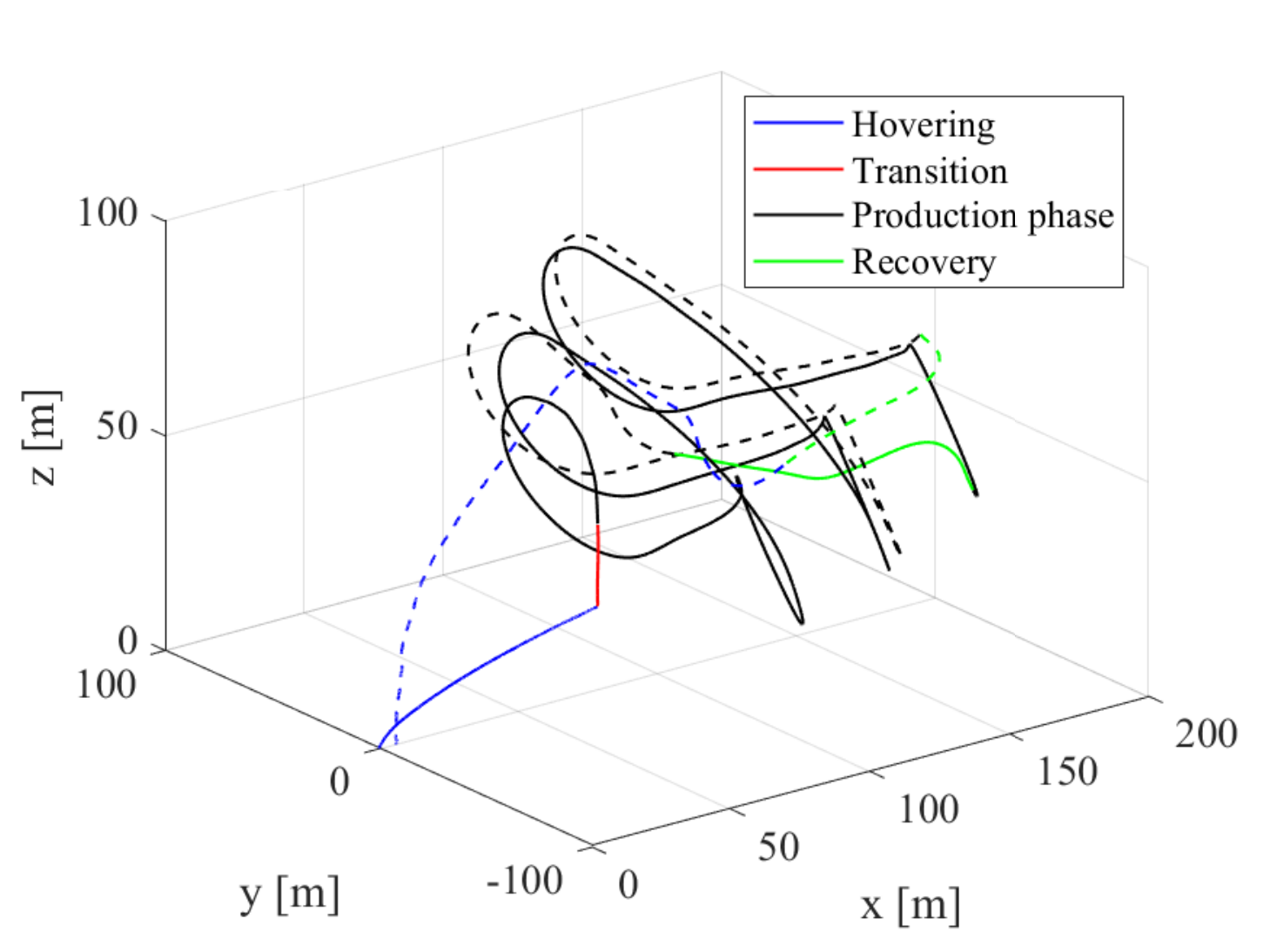}
\caption{Pumping operation with free-flight reentry: 3D drone path.}
\label{Path3d}
\end{figure}

\begin{figure}
\centering
\includegraphics[width=1\columnwidth, keepaspectratio]{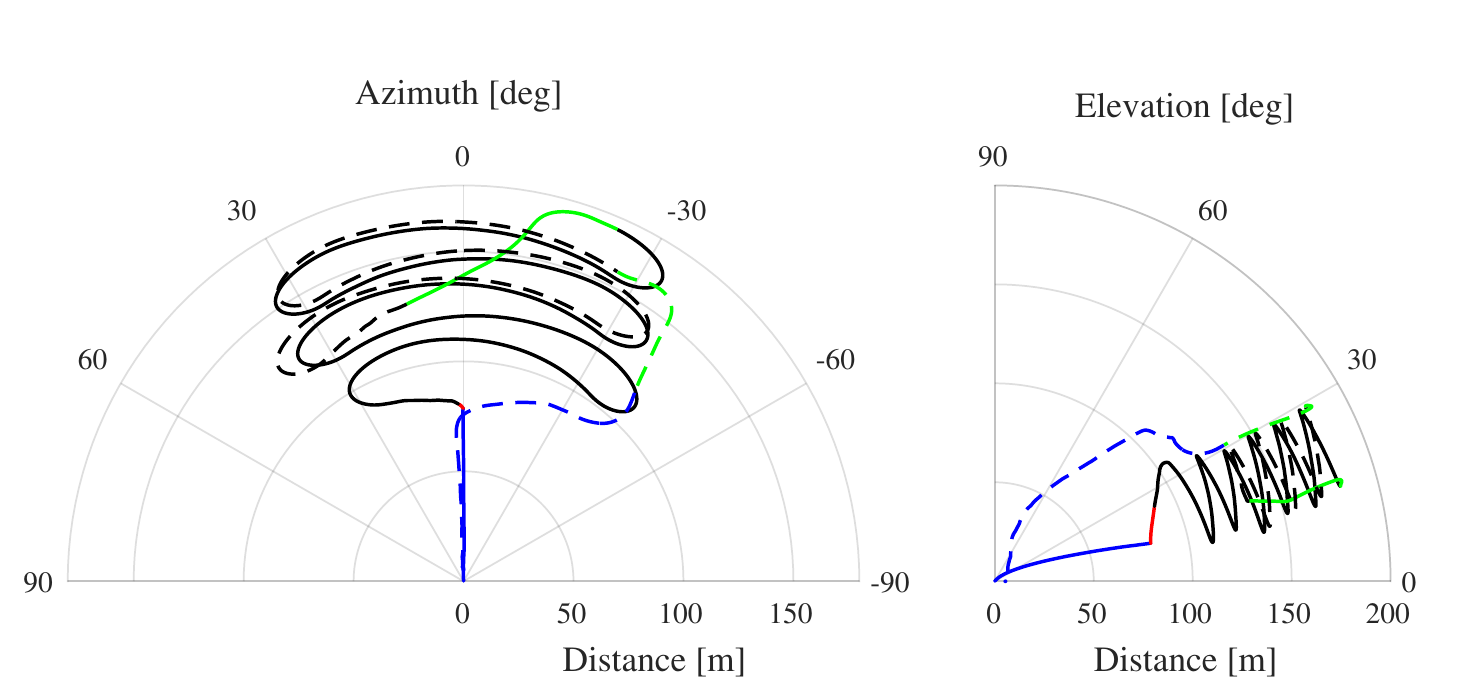}
\caption{Pumping operation with free-flight reentry: path in polar coordinates.}
\label{freeReentry_coord_unique}
\end{figure}

\begin{figure}
\centering
\includegraphics[width=1\columnwidth, keepaspectratio]{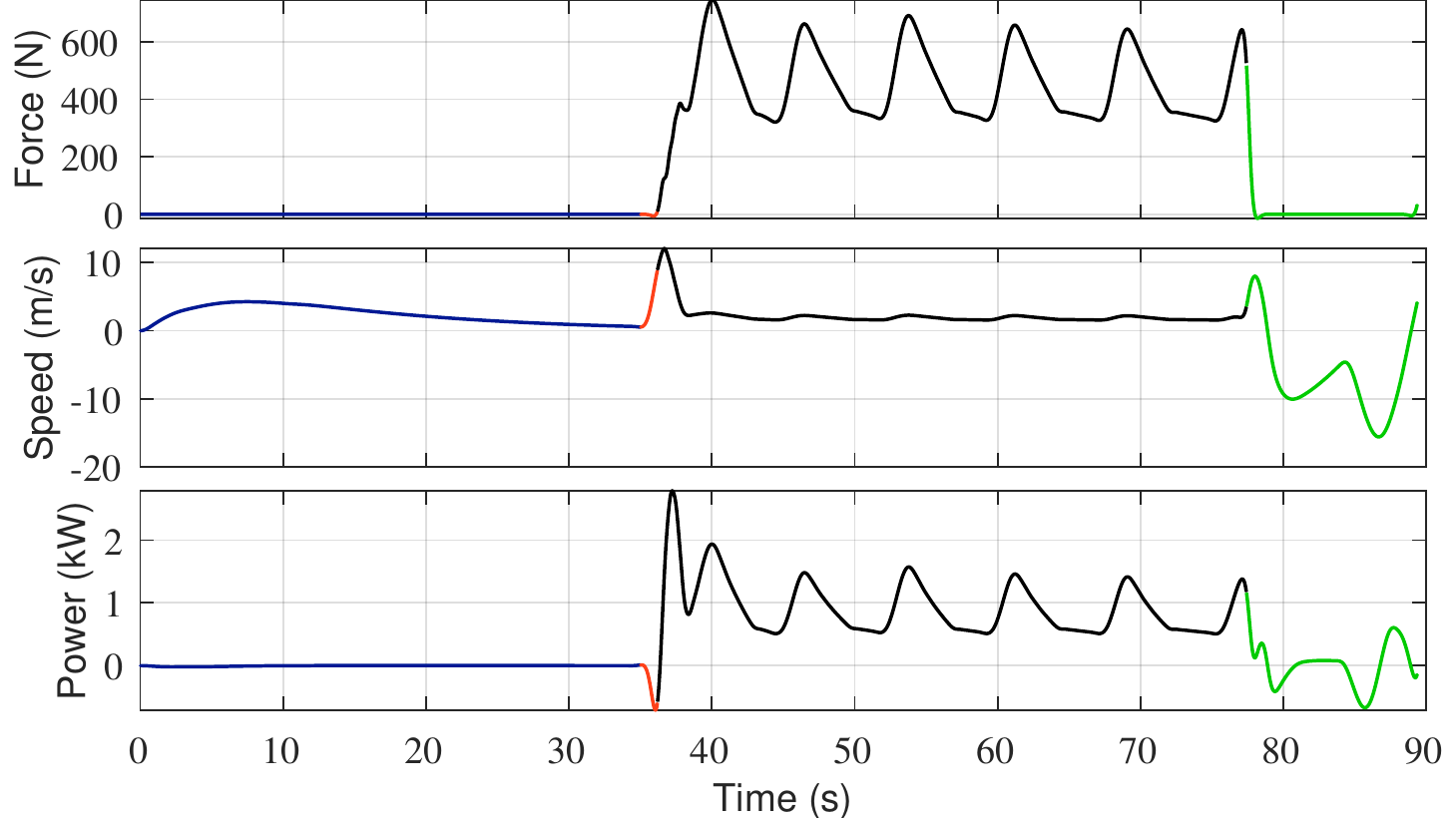}
\caption{Pumping operation with free-flight reentry: ground station variables.}
\label{GS_variables}
\end{figure}

\begin{figure}
\centering
\includegraphics[width=1\columnwidth, keepaspectratio]{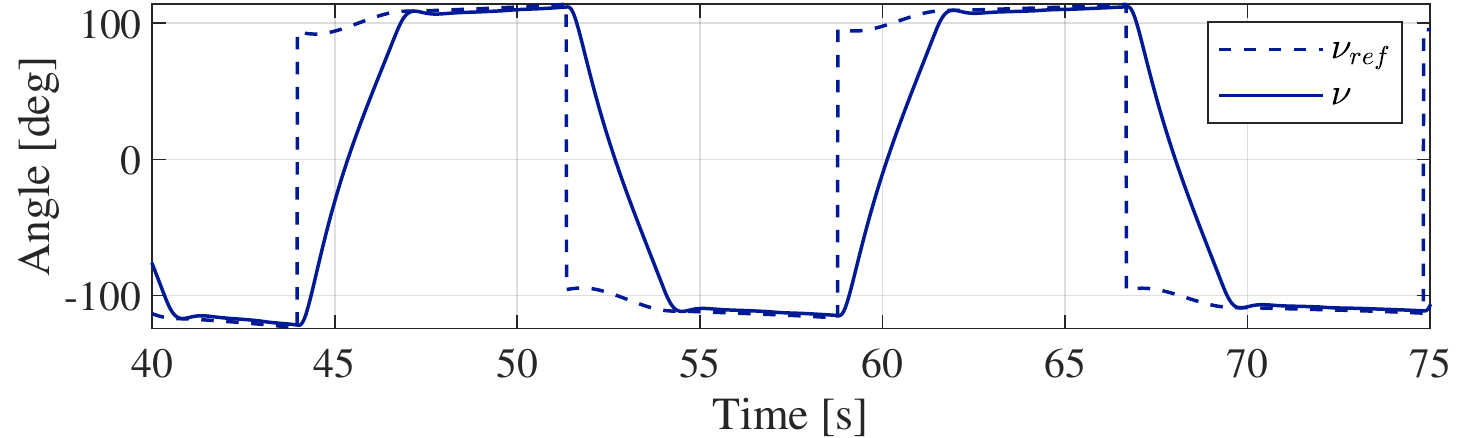}
\caption{Production phase: velocity angle tracking during figure-of-eight path.}
\label{v_angle}
\end{figure}

Several full flight simulations with tether and ground station have been performed. The relevant system parameters are shown in Table \ref{Tab:Parameters}, and the aerodynamic coefficients as a function of $\alpha,\,\beta$ are presented in Figure \ref{F:aero_coeff}. Controller poles have been set according to Table \ref{controllers_poles}. Wind has been set at 7 m/s mean value in $x_F$ direction. As described in the previous sections, the drone is commanded to take-off from the ground station in hovering mode until it reaches the point designated to start of active phase. Then, transition from hovering to tethered flight starts: the drone shall rapidly increase its speed and start to turn to one side, while for ground station the generation phase controller is enabled. When the front speed is sufficiently high, the tethered controller is enabled and the drone starts the generation phase, tracking figure of eight paths. When the length of the tether reaches a certain value, the reentry phase begin. We comment next the behavior when a free-flight reentry is implemented, since it results to be the most efficient strategy. In section \ref{SS:cycle}, we compare this approach with the other two alternatives described in Section \ref{SS:cycle}. When the reentry phase is concluded (the tether length is back to the starting value), another pumping cycle starts. Figure \ref{Path3d} shows the path performed by the drone: after the first hovering phase, two cycles with downloops are done. Figure \ref{freeReentry_coord_unique} shows the same path in terms of spherical coordinates. In the first one, it can be seen that the azimuth angle is kept between -40 and +40 degrees. In the second, it can be seen that the elevation angle during active phase is between 30 and 15 degrees. 
 The tracking of reference points is performed by means of the velocity angle (Figure \ref{v_angle}). 
In Figure \ref{GS_variables}, the main tether's quantities are shown: the tether force has a variation of more than 200 N during active phase with a mean value around 500 N, and  the tether is released at around 2 m/s. The mean power generated during 2 cycles is 541 W, while considering only the active phase the mean power generated is 908.6 W.

\subsection{Reentry phase comparison}
\begin{figure}
\centering
\includegraphics[width=1\columnwidth, keepaspectratio]{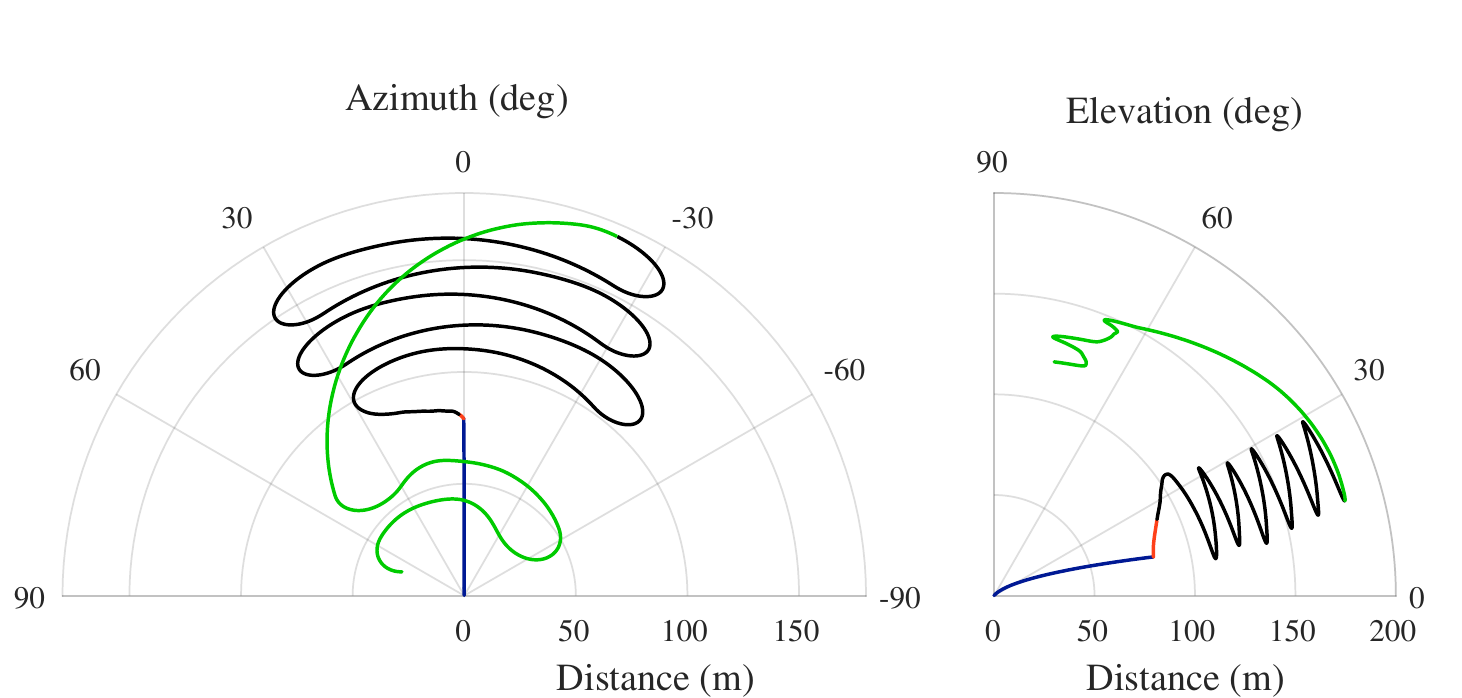}
\caption{Climb and descend reentry strategy:  elevation in polar plot.}
\label{HighEl_polar}
\end{figure}

We compared the three reentry phase strategies introduced in \ref{SS:cycle} in simulation, in order to study the advantages and drawbacks of each one. From the point of view of implementation, the simplest strategies are those with taut tether, because they avoid transitions between taut and slack tether, decreasing the number of involved controllers and the complexity of coordinating them.\\
Figure \ref{HighEl_polar} shows the trajectory in terms of azimuth and elevation angles for the climb and descend reentry strategy. The drone continues to perform figure-of-eight paths but at higher elevation angle, above 60 degrees, thus reducing the apparent wind seen by the tether. In this way, the tether can be reeled in under relatively lower force.

\begin{figure}
\centering
\includegraphics[width=1\columnwidth, keepaspectratio]{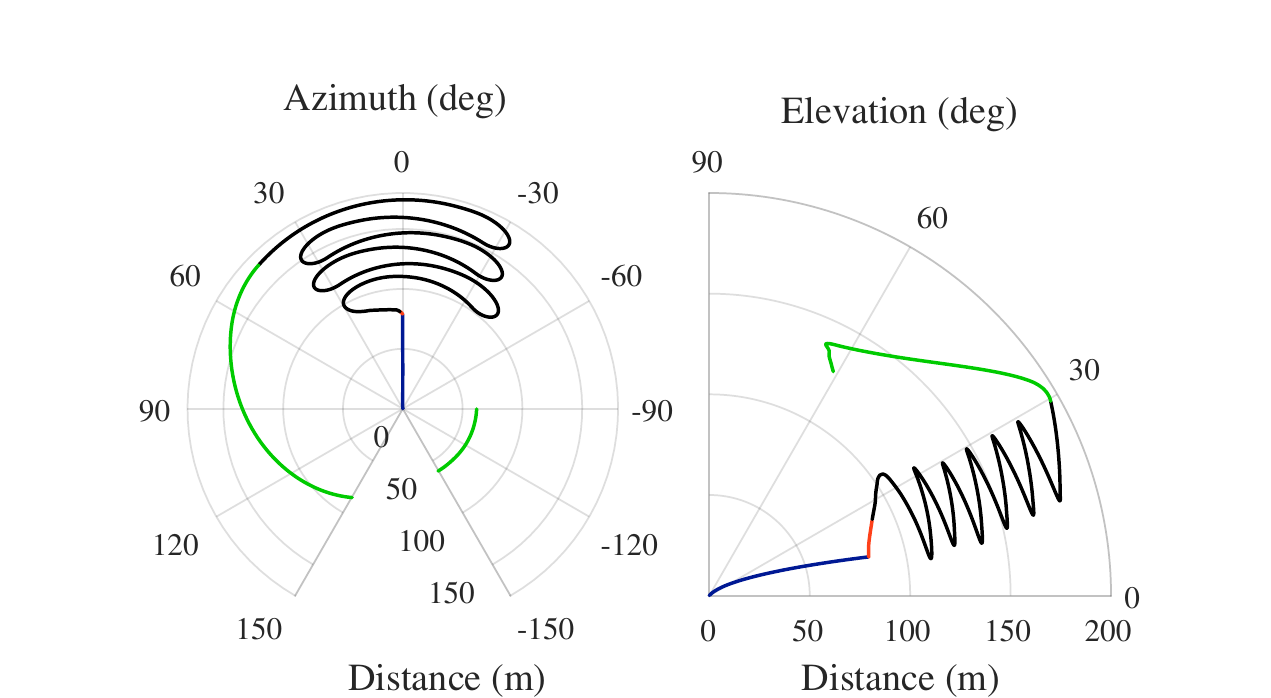}
\caption{Rotation around ground station reentry strategy:  elevation in polar plot.}
\label{BehindGSEl_polar}
\end{figure}

\begin{figure}
\centering
\includegraphics[width=1\columnwidth, keepaspectratio]{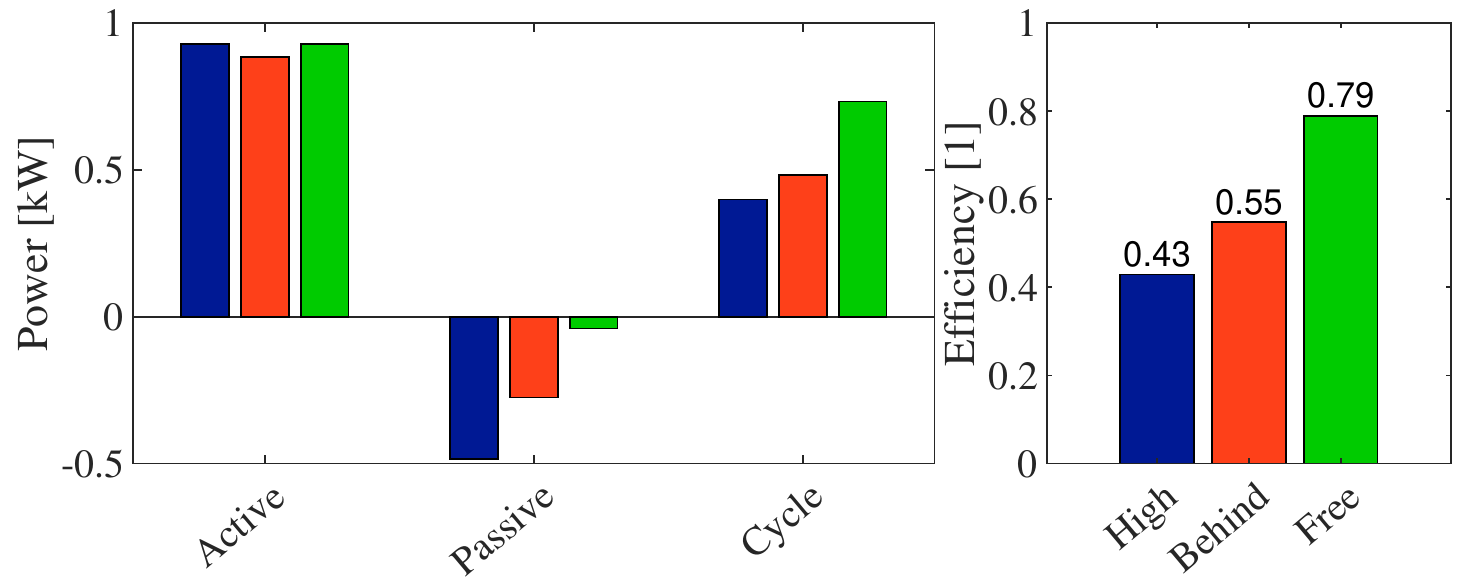}
\caption{Comparison among reentry strategies. Left: mean power in generation during the traction phase, mean power consumed in the retraction phase, and average cycle power. Right: cycle efficiency.}
\label{F:BarPlot}
\end{figure}

Figure \ref{BehindGSEl_polar} shows the trajectory with the reentry strategy consisting of a complete rotation around the ground station. In the first part of the reentry phase, the apparent wind is very high, because the drone is flying downwind, while when the drone moves upwind with respect to the ground station the apparent wind is very low, and the tether force decreases substantially.\\
%
In Figure \ref{F:BarPlot} the three strategies are compared in terms of cycle efficiency. The best reentry strategy in terms of efficiency is clearly the free-flight one, because it is fast and requires very low power. Instead, comparing the strategies with taut tether, it turns out that the  complete rotation around the ground station has a better efficiency (about 0.55) than the climb and descend strategy. This is due to the fact that it is easier to pull the tether when the drone is in the upwind zone. The climb and descend reentry strategy has the lowest efficiency among the three, however the average cycle power is still positive.

\section{Conclusions and suggested next steps}\label{S:conclusions}
The models and controllers described in this work form a complete simulation suite that can be implemented in numerical ODE solvers to evaluate the system behavior and to support experimental testing and product development for VTOL pumping AWE systems in all operating conditions. Through this model, a comparison between different reentry strategies has been carried out. The results show that the reentry strategy in free flight is the best solution in terms of efficiency but requires very accurate positioning and coordination between drone and ground station. Other, less complex solutions are available at the cost of lower cycle efficiency. \\
In addition to this, a steering authority analysis has been conducted, with informative results pertaining to the design phase of the aircraft.\\ 
The natural next step of this research is to test the control approach on a real prototype, in order to validate and improve the simulation environment.


\begin{thebibliography}{10}
	\providecommand{\url}[1]{#1}
	\csname url@samestyle\endcsname
	\providecommand{\newblock}{\relax}
	\providecommand{\bibinfo}[2]{#2}
	\providecommand{\BIBentrySTDinterwordspacing}{\spaceskip=0pt\relax}
	\providecommand{\BIBentryALTinterwordstretchfactor}{4}
	\providecommand{\BIBentryALTinterwordspacing}{\spaceskip=\fontdimen2\font plus
		\BIBentryALTinterwordstretchfactor\fontdimen3\font minus
		\fontdimen4\font\relax}
	\providecommand{\BIBforeignlanguage}[2]{{%
			\expandafter\ifx\csname l@#1\endcsname\relax
			\typeout{** WARNING: IEEEtran.bst: No hyphenation pattern has been}%
			\typeout{** loaded for the language `#1'. Using the pattern for}%
			\typeout{** the default language instead.}%
			\else
			\language=\csname l@#1\endcsname
			\fi
			#2}}
	\providecommand{\BIBdecl}{\relax}
	\BIBdecl
	
	\bibitem{Cherubini2015}
	A.~Cherubini, A.~Papini, R.~Vertechy, and M.~Fontana, ``Airborne wind energy
	systems: A review of the technologies,'' \emph{Renewable and Sustainable
		Energy Reviews}, vol.~51, pp. 1461--1476, 2015.
	
	\bibitem{Schmehl2018}
	R.~Schmehl, Ed., \emph{Airborne Wind Energy - Advances in Technology
		Development and Research}.\hskip 1em plus 0.5em minus 0.4em\relax Singapore:
	Springer, 2018.
	
	\bibitem{IRENA16}
	\BIBentryALTinterwordspacing
	{International Renewable Energy Agency (IRENA), Abu Dhabi, ``Innovation
		Outlook: Off-shore Wind''}. accessed on 17/1/2019. [Online]. Available:
	\url{https://www.irena.org/publications/2016/Oct/Innovation-Outlook-Offshore-Wind}
	\BIBentrySTDinterwordspacing
	
	\bibitem{Study_AWES}
	\BIBentryALTinterwordspacing
	{European Commission - Directorate-General for Research and Innovation, ``Study
		on Challenges in the commercialisation of airborne wind energy systems'',
		September 2018. Last accessed: 17/1/2019}. [Online]. Available:
	\url{https://op.europa.eu/en/publication-detail/-/publication/a874f843-c137-11e8-9893-01aa75ed71a1/language-en/format-PDF/source-112940934}
	\BIBentrySTDinterwordspacing
	
	\bibitem{WATSON2019109270}
	\BIBentryALTinterwordspacing
	S.~W. et~al., ``Future emerging technologies in the wind power sector: A
	european perspective,'' \emph{Renewable and Sustainable Energy Reviews}, vol.
	113, p. 109270, 2019. [Online]. Available:
	\url{http://www.sciencedirect.com/science/article/pii/S1364032119304782}
	\BIBentrySTDinterwordspacing
	
	\bibitem{FaMP09}
	L.~Fagiano, M.~Milanese, and D.~Piga, ``High-altitude wind power generation,''
	\emph{IEEE Transactions on Energy Conversion}, vol.~25, no.~1, pp. 168 --180,
	mar. 2010.
	
	\bibitem{Archer2009}
	C.~L. Archer and K.~Caldeira, ``Global assessment of high-altitude wind
	power,'' \emph{Energies}, vol.~2, no.~2, pp. 307--319, 2009.
	
	\bibitem{Archer2014}
	C.~L. Archer, L.~D. Monache, and D.~L. Rife, ``Airborne wind energy: Optimal
	locations and variability,'' \emph{Renewable Energy}, vol.~64, pp. 180--186,
	2014.
	
	\bibitem{BECHTLE20191103}
	P.~Bechtle, M.~Schelbergen, R.~Schmehl, U.~Zillmann, and S.~Watson, ``Airborne
	wind energy resource analysis,'' \emph{Renewable Energy}, vol. 141, pp. 1103
	-- 1116, 2019.
	
	\bibitem{VanderLind2013}
	D.~{Vander Lind}, ``Analysis and flight test validation of high performance
	airborne wind turbines,'' in \emph{Airborne Wind Energy}.\hskip 1em plus
	0.5em minus 0.4em\relax Berlin Heidelberg: Springer, 2013, ch.~28, pp.
	473--490.
	
	\bibitem{BAUER2018290}
	\BIBentryALTinterwordspacing
	F.~Bauer, R.~M. Kennel, C.~M. Hackl, F.~Campagnolo, M.~Patt, and R.~Schmehl,
	``Drag power kite with very high lift coefficient,'' \emph{Renewable Energy},
	vol. 118, pp. 290 -- 305, 2018. [Online]. Available:
	\url{http://www.sciencedirect.com/science/article/pii/S0960148117310285}
	\BIBentrySTDinterwordspacing
	
	\bibitem{Erhard2015}
	M.~Erhard and H.~Strauch, ``Flight control of tethered kites in autonomous
	pumping cycles for airborne wind energy,'' \emph{Control Engineering
		Practice}, vol.~40, pp. 13--26, 2015.
	
	\bibitem{Zgraggen2016}
	A.~Zgraggen, L.~Fagiano, and M.~Morari, ``Automatic retraction and full-cycle
	operation for a class of airborne wind energy generators,'' \emph{IEEE
		Transactions on Control Systems Technology}, vol.~24, no.~2, pp. 594--698,
	2016.
	
	\bibitem{LICITRA201815}
	G.~Licitra, A.~Bürger, P.~Williams, R.~Ruiterkamp, and M.~Diehl, ``Optimal
	input design for autonomous aircraft,'' \emph{Control Engineering Practice},
	vol.~77, pp. 15 -- 27, 2018.
	
	\bibitem{LKBWRD19}
	\BIBentryALTinterwordspacing
	G.~Licitra, J.~Koenemann, A.~B\"{u}rger, P.~Williams, R.~Ruiterkamp, and
	M.~Diehl, ``{Performance assessment of a rigid wing Airborne Wind Energy
		pumping system},'' \emph{Energy}, vol. 173, no.~C, pp. 569--585, 2019.
	[Online]. Available:
	\url{https://ideas.repec.org/a/eee/energy/v173y2019icp569-585.html}
	\BIBentrySTDinterwordspacing
	
	\bibitem{Fechner2015}
	U.~Fechner, R.~{van der Vlugt}, E.~Schreuder, and R.~Schmehl, ``Dynamic model
	of a pumping kite power system,'' \emph{Renewable Energy}, 2015.
	
	\bibitem{Heilmann2013}
	J.~Heilmann and C.~Houle, ``Economics of pumping kite generators,'' in
	\emph{Airborne Wind Energy}, U.~Ahrens, M.~Diehl, and R.~Schmehl, Eds.\hskip
	1em plus 0.5em minus 0.4em\relax Berlin Heidelberg: Springer, 2013, ch.~15,
	pp. 271--284.
	
	\bibitem{ch26-RuSo14}
	R.~Ruiterkamp and S.~Sieberling, \emph{Airborne Wind Energy}, ser. Green Energy
	and Technology.\hskip 1em plus 0.5em minus 0.4em\relax Berlin:
	Springer-Verlag, 2014, ch. 26. Description and Preliminary Test Results of a
	Six Degrees of Freedom Rigid Wing Pumping System, p. 443.
	
	\bibitem{Stuyts2015}
	J.~Stuyts, G.~Horn, W.~Vandermeulen, J.~Driesen, and M.~Diehl, ``Effect of the
	electrical energy conversion on optimal cycles for pumping airborne wind
	energy,'' \emph{IEEE Transactions on Sustainable Energy}, vol.~6, no.~1, pp.
	2--10, 2015.
	
	\bibitem{FNRSO18}
	L.~Fagiano, E.~Nguyen-Van, F.~Rager, S.~Schnez, and C.~Ohler, ``Autonomous take
	off and flight of a tethered aircraft for airborne wind energy,'' \emph{IEEE
		transactions on control systems technology}, vol.~26, no.~1, pp. 151--166,
	2018.
	
	\bibitem{VCSPW18}
	\BIBentryALTinterwordspacing
	K.~Vimalakanthan, M.~Caboni, J.~Schepers, E.~Pechenik, and P.~Williams,
	``Aerodynamic analysis of ampyx's airborne wind energy system,''
	\emph{Journal of Physics: Conference Series}, vol. 1037, p. 062008, jun 2018.
	[Online]. Available:
	\url{https://doi.org/10.1088%2F1742-6596%2F1037%2F6%2F062008}
		\BIBentrySTDinterwordspacing
		
		\bibitem{Twingtec2018}
		R.~Luchsinger, D.~Aregger, F.~B.~D. Costa, C.~Galliot, F.~Gohl, J.~Heilmann,
		H.~Hesse, C.~Houle, T.~A. Wood, and R.~S. Smith, \emph{Pumping Cycle Kite
			Power with Twings, in Schmehl R. (eds). Airborne Wind Energy. Green Energy
			and Technology}.\hskip 1em plus 0.5em minus 0.4em\relax Singapore: Springer,
		2018.
		
		\bibitem{Todeschini2019}
		D.~Todeschini, L.~Fagiano, C.~Micheli, and A.~Cattano, ``Control of vertical
		take off, dynamic flight and landing of hybrid drones for airborne wind
		energy systems,'' in \emph{2019 American Control Conference (ACC)}.\hskip 1em
		plus 0.5em minus 0.4em\relax IEEE, 2019, pp. 2177--2182.
		
		\bibitem{RSOH19}
		S.~Rapp, R.~Schmehl, E.~Oland, and T.~Haas, ``Cascaded pumping cycle control
		for rigid wing airborne wind energy systems,'' \emph{Journal of Guidance,
			Control, and Dynamics}, vol.~42, no.~11, pp. 2456--2473, 2019.
		
		\bibitem{HaOD18}
		H.~Li, D.~Olinger, and M.~Demetriou, \emph{Attitude Tracking Control of an
			Airborne Wind Energy System}, 04 2018, pp. 215--239.
		
		\bibitem{Bont10}
		E.~Bontekoe, ``Up! - how to launch and retrieve a tethered aircraft,'' Master's
		thesis, TU Delft, August 2010, accessed in September 2018 at
		http://repository.tudelft.nl/.
		
		\bibitem{Fagianoa}
		L.~Fagiano and S.~Schnez, ``On the take-off of airborne wind energy systems
		based on rigid wings,'' \emph{Renewable Energy}, vol. 107, pp. 473--488,
		2015.
		
		\bibitem{Skypull}
		\BIBentryALTinterwordspacing
		{Skypull SA}. [Online]. Available: \url{https://www.skypull.technology/}
		\BIBentrySTDinterwordspacing
		
		\bibitem{Loyd1980}
		M.~L. Loyd, ``{Crosswind kite power},'' \emph{Journal of Energy}, vol.~4,
		no.~3, pp. 106--111, 1980.
		
		\bibitem{FaMi12}
		L.~Fagiano and M.~Milanese, ``Airborne wind energy: an overview,'' in
		\emph{American Control Conference 2012}, Montreal, Canada, 2012, pp.
		3132--3143.
		
		\bibitem{etkin}
		B.~Etkin, \emph{Dynamics of {Atmospheric} {Flight}}.\hskip 1em plus 0.5em minus
		0.4em\relax Dover, 2005.
		
		\bibitem{Fagiano2009}
		\BIBentryALTinterwordspacing
		L.~Fagiano, ``Control of tethered airfoils for high-altitude wind energy
		generation,'' Ph.D. dissertation, Politecnico di Torino, 2009. [Online].
		Available:
		\url{https://fagiano.faculty.polimi.it/docs/PhD_thesis_Fagiano_Final.pdf}
		\BIBentrySTDinterwordspacing
		
		\bibitem{FZMK14}
		L.~Fagiano, A.~Zgraggen, M.~Morari, and M.~Khammash, ``Automatic crosswind
		flight of tethered wings for airborne wind energy: modeling, control design
		and experimental results,'' \emph{IEEE Transactions on Control Systems
			Technology}, vol.~22, no.~4, pp. 1433--1447, 2014.
		
	\end{thebibliography}
\end{document}